\documentclass[twocolumn,useamsfonts,mfastrosym,usegraphicx,useepstopdf]{pasj01}
\usepackage{color}
\usepackage{graphicx,wrapfig}
\usepackage{comment}
\usepackage{hhline}
\usepackage{CJK}
\usepackage{morefloats}
\usepackage{graphicx}
\usepackage[toc]{appendix}
\usepackage{natbib}
\usepackage{epstopdf}  % for pdflatex

\def\mbf#1{\mbox{\boldmath ${#1}$}}
\def\Alfven{Alfv\'{e}n~}
\def\Alfvenic{Alfv\'{e}nic~}

\begin{document}
%\begin{CJK*}{UTF8}{gbsn}

\title{Stellar Winds and Coronae \\of Low-mass Pop. II/III Stars}

%\author{Takeru K. Suzuki (鈴木 建)$^{1}$ \& Shu-ichiro Inutsuka (犬塚 修一郎)$^{1}$}
\author{Takeru K. Suzuki$^{1,2}$}
\email{stakeru@ea.c.u-tokyo.ac.jp}
\altaffiltext{1}{School of Arts \& Sciences, University of Tokyo,
3-8-1, Komaba, Meguro, Tokyo 153-8902, Japan ; }
\altaffiltext{2}{
Department of Physics, Nagoya University, Furo-cho, Chikusa,
Nagoya, Aichi 464-8602, Japan %; %}
%\altaffiltext{3}{
%Department of Astronomy, the University of Tokyo,
%7-3-1, Hongo, Bunkyo, Tokyo 113-8654, Japan
}
\SetRunningHead{Suzuki}{Winds from Low-mass Pop. II/III Stars}


\maketitle

\begin{abstract}
  We investigated stellar winds from zero/low-metallicity low-mass stars
  by magnetohydrodynamical simulations
  for stellar winds driven by \Alfven waves from stars with mass $M_{\star}
  =(0.6-0.8)M_{\odot}$ and metallicity $Z=(0-1)Z_{\odot}$, where $M_{\odot}$
  and $Z_{\odot}$ are the solar mass and metallicity, respectively.
%  We inject velocity fluctuation by the surface convection from the photosphere.
  \Alfvenic waves, which are excited by the surface convection, travel
  upward from the photosphere and heat up the corona by their dissipation.
  For lower $Z$, denser gas can be heated up to the coronal temperature
  because of the inefficient radiation cooling.  
  The coronal density of Pop.II/III stars with $Z\le 0.01Z_{\odot}$ is 1-2 orders
  of magnitude larger than that of the solar-metallicity star with the same
  mass, and as a result, the mass loss rate, $\dot{M}$,
  %is also larger for stars with lower
%  metallicity, which is a direct outcome of the larger coronal density;
%  $\dot{M}$ of the Pop.II/III stars
  is $(4.5-20)$ times larger.
  This indicates that metal accretion on low-mass Pop.III stars is negligible.
  %  than $\dot{M}$ of the solar metallicity star with the same mass.
  The soft X-ray flux %[erg cm$^{-2}$s$^{-1}$]
  of the Pop.II/III stars is also
  expected to be $\approx (1-30)$ times larger than that of the solar-metallicity
  counterpart owing to the larger coronal density, even though the radiation
  cooling efficiency %[erg cm$^{3}$s$^{-1}$]
  is smaller.
%  The larger coronal density also leads to a larger transmissivity of the
%  \Alfvenic waves to the corona, because they propagate a shorter distance
%  with the smaller density difference from the photosphere to the corona and
%  can avoid severe reflection. 
  %, which is a universal feature independent from metallicity.
  A larger fraction of the input \Alfvenic wave energy is transmitted to
  the corona in low $Z$ stars because they avoid severe reflection owing to
  the smaller density difference between the photosphere and the corona.
  Therefore, a larger fraction is converted to the thermal
  energy of the corona and the kinetic energy of the stellar wind.
  %because of the
  %suppression of the radiation cooling in the lower atmospheric region.
  From this energetics argument, we finally derived a scaling of $\dot{M}$ as 
  $\dot{M}\propto L R_{\star}^{11/9}M_{\star}^{-10/9}T_{\rm eff}^{11/2}\left[\max
    (Z/Z_{\odot},0.01)\right]^{-1/5}$, where $L$, $R_{\star}$, and $T_{\rm eff}$
  are stellar luminosity,  radius, and effective temperature, respectively.
%  In addition to this, stars with lower-metallicity stars are expected to have
%  larger convective velocity because of the larger surface
%  temperature, which also contributes to the larger $\dot{M}$ for lower-
%  metallicity stars.  
\end{abstract}

\section{Introduction}
\label{sec:intro}
Various kinds of stars, and probably all the stars, drive stellar winds from
their surfaces. Radiation pressure plays a major role in stellar winds from
luminous stars; in massive stars located in the bluer side of a
Hertzsprung-Russell (HR) diagram, the stellar winds are accelerated by the
absorption of the ultraviolet (UV) radiation on metallic lines, which are called
line-driven winds \citep[][]{ls70,cak75}; in asymptotic giant branch (AGB)
stars located in the redder side of a HR diagram, the absorption of the
infrared (IR) radiation by dust grains is believed to be the main driver
of the winds \citep{bow88,fh08,ohn16,ohn17}.  
Since heavy elements play an essential role in these types of stellar winds,
the mass loss rate, $\dot{M}$, of line-driven winds
\citep{kud02,mui12} and dust-driven winds \citep{wac08,tas17} are positively
correlated with metallicity. 

On the other hand, in less luminous stars, radiation pressure cannot be the
leading part to drive stellar winds; instead, magnetohydrodynamical
(MHD hereafter) processes play a major role.
Low-mass main sequence stars with the stellar mass
$M_{\star} \lesssim M_{\odot}$ have a surface convection zone,
which excites various types of waves. %, where $M_{\odot}$ is the solar mass.
Among various modes of waves, the \Alfven wave, which travels a long
distance to the upper atmosphere on account of the less dissipative character,
is believed to contribute to the acceleration of the stellar wind
%, and has been investigated
\citep{bel71,od98,si05,vv07,cra07,cs11,suz13}. 
This mechanism is also considered to operate up to moderately evolved red giant
stars \citep{air00,suz07,air10}.
\citet{cs11} derived mass loss rates of these types of stars by time-steady
calculations with taking into account the effect on metallicity, whereas
the explicit dependence of wind properties, e.g., $\dot{M}$, on metallicity
was not presented.
%, probably because their main aim was direct comparison to observed data in wide ranges of other stellar parameters.

The main purpose of the present paper is to investigate the dependence of the
mass loss rate and atmospheric properties of low-mass low-metallicity stars
on metallicity by time-dependent MHD simulations.
Mass loss rates of low-mass stars with [Fe/H]$<-1$ and $M_{\star}\le M_{\odot}$
have not been observationally obtained to date.
Therefore, our results for low-metallicity stars
%are based on purely theoretical interests
cannot be directly compared to observational data at present.
However, these stars could be a direct link between the present-day universe
and early epochs during the structure formation was going on. 
%very important in terms of observational cosmology and structure formation. 

The formation of first stars, which are called Population III (Pop.III
hereafter) stars, have been paid much attention.
It has been argued that massive stars are favorably formed
in metal-free circumstances, because the Jeans mass is larger owing
to the inefficient cooling \citep[e.g.,][]{on98,bro02,abe02,omu05,yos06,yos08,hos13,fuk18}.
However, %once after massive first stars form,
recent studies show that low-mass metal-free stars are also possibly formed 
%Strong ultraviolet (UV hereafter) radiation of the massive Pop.III stars reduces the Jeans mass (REFs).
through fragmentation %in collapsing clouds
in accreting protostellar disks around primary massive proto-Pop.III
stars \citep{mac08,cla11,gre11,md13,sus14,chi16}.

If such low-mass Pop.III stars with $M_{\star}\lesssim 0.8M_{\odot}$ are really
formed, we can directly observe them in the present universe, because their
lifetimes are longer than the age of the universe
\citep[$=13.8$ Gyr;][]{pla16}. Although a large number
of low-mass metal-poor stars have been detected \citep[e.g.,][and references
  therein]{aok06,fre15}, a low-mass
zero-metal star has not been identified to date. A possible interpretation
of the non-detection is accretion of heavy elements; even though a star is
purely metal-free at the formation, the surface is gradually polluted with time
via traveling through the interstellar medium \citep{yos81,kom15,she17}.

However, \citet{tan17} recently pointed out that stellar wind from low-mass
Pop.III stars can almost block accreting gas and the pollution is negligible
if the wind flux is comparable to that of the solar wind.  
The amount of accreting material depends on the properties of the stellar
winds. Determining the mass flux and velocity of winds from low-mass
Pop.III stars, 
%by MHD simulation
which is one of the main purposes of this paper, is crucial to evaluate
this surface pollution mechanism in a quantitative manner. 

The construction of the paper is as follows. 
In Section \ref{sec:setup} we describe our MHD simulations. 
We present main results of the MHD simulations in Section \ref{sec:res} and
discuss related topics and limitations of our treatment in Section
\ref{sec:dis}. We summarize the paper in Section \ref{sec:sum}.

\section{Setup}
\label{sec:setup}
\begin{table*}
  {\small 
  \begin{tabular}{|c|c|c|c|c|c|c|c|c|c||c|c|}
    \hline
    $M_{\star} [M_{\odot}]$ & $Z [Z_{\odot}]$ & $R_{\star} [R_{\odot}]$
    & $T_{\rm eff}[K]$ & $L [L_{\odot}]$ & $\rho_{{\rm ph},7}$
    & $\delta v_0${\footnotesize [km s$^{-1}$]} & $B_{r,0}$[kG] & $f_0$
    & $h_{\rm l}${\footnotesize [$0.01R_{\star}$]}
    & $\dot{M}_{14}$ & $v_{\rm t}${\footnotesize [km s$^{-1}$]}\\
    \hline    
    \hline
    0.8 & 1 & 0.737 & 5096 & 0.328 & 4.37 & 0.877 & 1.96 & 1/1570 & 1.62 & 0.377 & 902\\
    \hline
    0.8 & 0.1 & 0.766 & 6030 & 0.695 & 2.85 & 1.27 & 1.72 & 1/1379 & 2.00 & 4.66 & 563 \\
    \hline
    0.8 & 0.01 & 0.771 & 6319 & 0.849 & 4.35 & 1.17 & 2.18 & 1/1744 & 2.11 & 7.70 & 433 \\
    \hline
    0.8 & 0 & 0.766 & 6365 & 0.863 & 5.00 & 1.13 & 2.35 & 1/1877 & 2.11 & 7.37 & 432 \\
    \hline
    \hline
    0.7 & 1 & 0.632 & 4657 & 0.169 & 7.80 & 0.641 & 2.51 & 1/2006 & 1.46 & 0.197 & 784 \\
    \hline
    0.7 & 0.1 & 0.620 & 5576 & 0.333 & 9.64 & 0.760 & 3.05 & 1/2441 & 1.71 & 0.470 & 793 \\
    \hline
    0.7 & 0.01 & 0.618 & 5815 & 0.391 & 9.32 & 0.812 & 3.06 & 1/2451 & 1.78 & 1.98 & 622 \\
    \hline
    0.7 & 0 & 0.617 & 5842 & 0.397 & 10.4 & 0.787 & 3.25 & 1/2600 & 1.78 & 1.95 & 608 \\
    \hline
    \hline
    0.6 & 1 & 0.546 & 4214 & 0.0842 & 10.7 & 0.505 & 2.80 & 1/2237 & 1.33 & 0.0783 & 890 \\
    \hline
    0.6 & 0.1 & 0.531 & 4976 & 0.155 & 12.8 & 0.594 & 3.32 & 1/2655 & 1.53 & 0.104 & 988\\
    \hline
    0.6 & 0.01 & 0.508 & 5303 & 0.183 & 19.6 & 0.561 & 4.24 & 1/3391 & 1.55 & 0.351 & 859\\
    \hline
    0.6 & 0 & 0.504 & 5344 & 0.186 & 23.5 & 0.533 & 4.67 & 1/3733 & 1.55 & 0.334 & 843 \\
    \hline
%    \hline
%    0.5 & 1 & 0.481 & 3645 & 0.0367 & 8.45 & 0.450 & 2.31 & 1/1848 & 1.21 & 0.0338 & 909\\
%    \hline
%    0.5 & 0.1 & 0.459 & 4306 & 0.0648 & 29.0 & 0.373 & 4.65 & 1/3718 & 1.37 & 0.0319 & 1070 \\
%    \hline
%    0.5 & 0.01 & 0.424 & 4636 & 0.0743 & 41.6 & 0.365 & 5.78 & 1/4625 & 1.36 & 0.0155 & 1170 \\
%    \hline
%    0.5 & 0 & 0.416 & 4803 & 0.0826 & 60.1 & 0.339 & 7.07 & 1/5656 & 1.38 & 0.0253 & 1150 \\
\hline
    1  & 1 & 1 & 5780 & 1 & 2.51 & 1.25 & 1.58 & 1/1265 & 2.00 & 2.22 & 690\\
    \hline
%    \hline
  \end{tabular}
  }
  \caption{Input parameters (1st -- 10th columns) and output properties
    (11th -- 12th columns) of stars with different masses and metallicities.
    $\rho_{{\rm ph},7}$ is the photospheric density at $T=T_{\rm eff}$ that
    is normalized by $10^{-7}$g cm$^{-3}$. $\dot{M}_{14}$ is time-averaged mass
    loss rate normalized by $10^{-14}M_{\odot}$ yr$^{-1}$.     
  \label{tab:stars_in}}
\end{table*}

We extended a MHD simulation code that was originally developed for the
solar wind \citep{si05,si06} to simulate stellar winds from
low-metallicity and low-mass stars.
Input parameters of our simulations are the strength and configuration of
magnetic field and velocity perturbation at the photosphere.
These parameters, which are essentially determined by the dynamo activity
in the surface convective layer, control heating the atmosphere and driving
stellar winds.
In our setup, we scale the input parameters from standard values calibrated by
the Sun. %current solar corona and wind. 

\subsection{A Standard Case for the Sun}
\label{sec:ssm}
We briefly explain a standard model for the Sun that is used for the scalings
of the input parameters of low/zero-metal stars. 
We slightly modified the basic setups of our previous simulations
\citep{si05,si06,suz13}. % but carried out modification. 
The main change %from the previous simulations
is that we set the inner boundary
at the location which the temperature coincides with the effective temperature,
$T_{\rm eff,\odot}=5780$ K. We determined the density, $\rho_{\rm ph,\odot}
=2.5\times 10^{-7}$g cm$^{-3}$, at this inner boundary from the ATLAS model
atmosphere \citep{kur79,ck03} for the Sun. We note that the Rosseland-mean
optical depth at this location is $0.37$ and that $\rho_{\rm ph,\odot}$ is larger
than the density $=10^{-7}$g cm$^{-3}$ at the inner boundary adopted
in our previous simulations \citep{si05,si06,suz13}.
%, which is located at an upper level in the photosphere.

\subsubsection{Magnetic Field}

We treated the solar wind in a magnetic flux tube that is rooted from a
kilogauss (kG) patch \citep{tsu08,st09,ito10,shi12} and superradially open
to the interplanetary space \citep{kh76}.
We assumed the equipartition between the magnetic pressure and the gas pressure
at the inner boundary,
\begin{equation}
  \frac{8\pi p_{0}}{B_{r,0}^2} = 1.
    \label{eq:eqpph}
\end{equation}
The gas pressure, $p_{0}$, at the inner boundary was determined from
$\rho_{\rm ph}$ and $T_{\rm eff}$ via an equation of state of ideal gas,
\begin{equation}
  p_0 = (\rho_{\rm ph}/\mu m_{\rm u}) k_{\rm B} T_{\rm eff},
\end{equation}
where $\mu$ is mean molecular weight, $m_{\rm u}$ is the atomic mass unit,
and $k_{\rm B}$ is the Boltzmann constant.
We here adopted $\mu=1.3$ as a standard value at the solar photosphere. 
Equation (\ref{eq:eqpph}) determines the magnetic field strength
$B_{r,0,\odot}=1.58$ kG at the inner boundary, which is a reasonable value for
typical kG-patches. 
%as shown in Table \ref{tab:stars_in}.

We fixed a super-radially open magnetic flux tube that is rooted from this
kG-patch. We basically followed a prescription of a super-radial expansion
factor introduced by \citet{kh76}, but redefined a filling factor of the open
flux tube regions, $f$, over the entire surface area,
$4\pi r^2$, at $r$ \citep{suz13},
\begin{equation}
f(r) = \frac{e^{\frac{r-R_{\star}-h_{\rm l}}{\zeta}} + f_0 - (1-f_0)e^{-\frac{h_{\rm l}}{\zeta}} }
{e^{\frac{r-R_{\star}-h_{\rm l}}{\zeta}}+1}, 
\label{eq:ff}
\end{equation}
where $h_{\rm l}$ corresponds to a typical height of closed loops and
$f_{0}(<1)$ is the filling factor at the stellar surface $r=R_{\star}$ ($R_{\star}
=R_{\odot}$ for the Sun).
Note that the super-radial expansion factor $=f(r)/f_0$.
%by \citet{kh76} is derived as
The flux tube expands most rapidly between $r=R_{\star}+h_{\rm l}-\zeta$ and
$r=R_{\star}+h_{\rm l}+\zeta$. We set $\zeta=\frac{1}{2}h_{\rm l}$
%which are tabulated in ... We set
and $h_{\rm l}=0.02R_{\odot}$ for our solar model. % \citep{suz13}. 
$f(r)\Rightarrow f_0(<1)$ for $r\Rightarrow R_{\star}$ and
$f(r)\Rightarrow 1$ for $r \Rightarrow \infty$.
%$f(r)$ is equivalent of a super-radial expansion factor of a flux tube
%introduced by \citet{kh76}, whereis it is usulally defined as
%$\rightarrow 1$ for $r\rightarrow R_{\star}$ and $\rightarrow 1/f_0(>1)$ for
%$\Rightarrow \infty$.
The profile of the radial component of the magnetic field is determined from
the adopted $f(r)$ by 
\begin{equation}
  B_r = B_{r,0}\frac{f_0 R_{\star}^2}{f(r)r^2}.
  \label{eq:divB0}
\end{equation}
%where $B_{r,{\rm ph}}$ is the strength of the radial magnetic field at
%the photosphere.
$f_0$ determines the average field strength of the open flux tube regions
at the photosphere, and we here adopted $f_0=1/1561$, which gives
\begin{equation}
B_{r,0}f_0 = 1.25\; {\rm G}. 
\label{eq:avB}
\end{equation}
The average field strength is stronger than this value, because
the contribution from closed magnetic loops is summed up to $B_{r,0}f_0$.
Recent observation by \citet{iid15} gives an average unsigned
magnetic flux density $\approx 2.5-4$ G in quiet-Sun regions, which
is moderately stronger than our adopted value and consistent with this
general picture.

\subsubsection{Velocity Perturbation}
\label{sec:dvsun}
We injected velocity perturbations from the inner boundary at the photosphere,
which excite MHD waves. 
We assumed the same amplitude for all the three (radial and transverse)
components at the photosphere.
We adopted the power spectrum, $P(\omega)\propto \omega^{-1}$, with frequency,
$\omega$, covering two orders of magnitude from $\omega_{\rm min}$ to
$\omega_{\rm max}=100\omega_{\rm min}$: 
\begin{equation}
  \langle \delta v_0^2\rangle = \int_{\omega_{\rm min}}^{\omega_{\rm max}} d\omega
  P(\omega),
  \label{eq:dv0}
\end{equation}
where we set $1/\omega_{\rm min} = 30$ min. and $1/\omega_{\rm max}=0.3$ min
in the standard case for the solar wind \citep{suz13}.
We adopted $\langle \delta v_{0,\odot}\rangle=1.25$ km s$^{-1}$ for the solar case, which
is consistent with observed velocity perturbation at the photosphere
$\approx 1.1$ km s$^{-1}$ \citep{mk10}, and well explains
the average properties of the solar wind (see later in this subsection
\ref{sec:ssm}). 

\subsubsection{MHD Code}
\label{sec:MHDc}
The velocity fluctuations injected from the photosphere
excite upgoing \Alfvenic (transverse) waves and
acoustic (longitudinal slow MHD) waves. We dynamically handled the
propagation, dissipation, and reflection of these waves.
We covered the simulation region from the photosphere to a sufficiently distant
location, $r_{\rm out}=30R_{\odot}$ ($\approx 0.15$ au), where $R_{\odot}$ is
the solar radius. A great advantage of our treatment is that we can
directly determine the mass loss rate from the surface convective perturbations. 
%Waves are treated in a 1D flux tube but
We took into account the three
components of magnetic and velocity field to handle \Alfvenic waves;
we time-dependently solved the following set of MHD equations with
radiative cooling, $q_{\rm R}$, and thermal conduction, $F_{\rm c}$
by 2nd order Godunov-MoC (Method-of-Characteristics)
method \citep{si05}: 
\begin{equation}
\label{eq:mass}
\frac{d\rho}{dt} + \frac{\rho}{r^2 f}\frac{\partial}{\partial r}
(r^2 f v_r ) = 0 , 
\end{equation}
\begin{displaymath}
\rho \frac{d v_r}{dt} = -\frac{\partial p}{\partial r}  
- \frac{1}{8\pi r^2 f}\frac{\partial}{\partial r}  (r^2 f B_{\perp}^2)
\end{displaymath}
\begin{equation}
\label{eq:mom}
+ \frac{\rho v_{\perp}^2}{2r^2 f}\frac{\partial }{\partial r} (r^2 f)
-\rho \frac{G M_{\star}}{r^2}  , 
\end{equation}
\begin{equation}
\label{eq:moc1}
\rho \frac{d}{dt}(r\sqrt{f} v_{\perp}) = \frac{B_r}{4 \pi} \frac{\partial} 
{\partial r} (r \sqrt{f} B_{\perp}).
\end{equation}
$$
\rho \frac{d}{dt}\left(e + \frac{v^2}{2} + \frac{B^2}{8\pi\rho}
- \frac{G M_{\star}}{r} \right) 
+ \frac{1}{r^2 f} 
\frac{\partial}{\partial r}\left[r^2 f \left\{ \left(p 
+ \frac{B^2}{8\pi}\right) v_r  \right. \right.
$$
\begin{equation}
\label{eq:eng}
\left. \left.
- \frac{B_r}{4\pi} (\mbf{B \cdot v})\right\}\right]
+  \frac{1}{r^2 f}\frac{\partial}{\partial r}(r^2 f F_{\rm c}) 
%+ \frac{q_{\rm R}}{\rho}=0 ,
+ q_{\rm R} = 0,
\end{equation}
\begin{equation}
\label{eq:ct}
\frac{\partial B_{\perp}}{\partial t} = \frac{1}{r \sqrt{f}}
\frac{\partial}{\partial r} [r \sqrt{f} (v_{\perp} B_r - v_r B_{\perp})], 
\end{equation}
where $G$, %$M_{\star}$,
$\rho$, $\mbf{v}$, $\mbf{B}$, $p$, and $e$ are
the gravitational constant, %stellar mass,
density, velocity, magnetic field, gas pressure, and internal energy,
respectively; $e$, $p$, and $\rho$ are related via
$e=\frac{p}{(\gamma-1)\rho}$ and we assumed the ratio of specific heats,
$\gamma=5/3$. The above equations are constructed in the spherical
coordinates, $(r,\theta,\phi)$, at $\theta=\pi/2$, and therefore we
did not distinguish $\theta$ and $\phi$ and simply used the subscript
$\perp$ ($=\theta$ and $\phi$). We note that the direction of $\theta=0$
has no relation with the magnetic or rotational axis of an actual star; our
magnetic flux tube can be located anywhere on the star.
%from polar to equatorial regions. 

We set up fine-scale grid points with spacing, $\Delta r<10$ km, from the
photosphere to the transition region, and gradually enlarged with r
to $\Delta r=2800$ km in the solar wind region according to the increase
of the \Alfven and sound velocities.
%The simulation region covers up to $r=r_{\rm out}=30R_{\odot}$ and
Above the outer boundary at $r_{\rm out}=30R_{\odot}$,
we prepared a buffer zone up to $200R_{\odot}$($\approx 1$ au), in which
$\Delta r$ increases to $\approx 10^6$km. At the outer boundary of the buffer
zone, we prescribed the outgoing boundary condition for both gas
and waves by using the seven characteristics of MHD waves \citep{si06}.

We injected the velocity perturbation from the photosphere
(see \S \ref{sec:dvsun}) but kept
the density ($=\rho_{\rm ph,\odot}$) and the temperature
($=T_{\rm eff,\odot}$) at the photosphere to the initial values 
%Table \ref{tab:stars_in}
throughout the simulation.
We also fixed $B_r$ in the entire
simulation region to the initial state throughout the simulation in order
to keep $\mbf{\nabla\cdot B}=0$ (equation \ref{eq:divB0}). 
Although we did not explicitly input perturbation of the magnetic field
from the photosphere, the transverse components, $\mbf{B}_{\perp}$,
were excited from one grid point above the photosphere by the injected
velocity perturbation. 

\subsubsection{Summary of the Solar Case}
The standard case for the Sun gave the time-averaged mass loss rate,
$\dot{M}=2.22\times 10^{-14}M_{\odot}$yr$^{-1}$, and terminal velocity,
$v_{\rm t}=690$ km s$^{-1}$, which explain the average properties of
the solar wind (Table \ref{tab:stars_in}). 

\subsection{Low/Zero-metallicity Stars}
\label{sec:LZMS}
Low-mass stars possess a surface convective zone, because the large opacity 
in the envelope region inhibits the effective energy transport by
radiation. Because the opacity has a positive dependence on
metallicity, the depth of the surface convection in lower-metallicity
stars is shallower.  
According to stellar evolution calculations by \citet{ric02a,ric02b},
a star with mass, $M_{\star}=0.9M_{\odot}$, and abundance of heavy elements,
$Z\lesssim 10^{-2}Z_{\odot}$, initially possesses a surface convective layer,
however, it shrinks with time and eventually disappears after $t>5$
Gyr before the end of the main sequence stage, where %$M_{\odot}$ and
$Z_{\odot}=0.014$ %are the solar mass and
is the solar metallicity 
%; we adopt $Z_{\odot}=0.014$ for
%the abundance of the sum of the heavy elements
%the elements heavier than Boron
\citep[][]{asp09}.

On the other hand, lower-mass stars with $M_{\star}\lesssim 0.85
M_{\odot}$ have a surface convection layer during the whole main
sequence duration. Since our focus is on MHD wave-driven stellar winds, of
which the original energy resides in the surface convection, we
considered low-mass stars with $M_{\star}\le 0.8 M_{\odot}$.
Table \ref{tab:stars_in} summarizes all the cases we simulated:
%cover in this paper,
stars with mass $M_{\star}=0.8M_{\odot},
0.7M_{\odot}, 0.6M_{\odot}$ and metallicity $Z=Z_{\odot}, 0.1Z_{\odot},
0.01Z_{\odot}, 0$. 
%We used the results of stellar evolution calculations by \citet{yi01,yi03}, 
%which we summarize %basic properties of the stars we calculate in and
We adopted 
basic stellar parameters, radius, $R_{\star}$, effective
temperature, $T_{\rm eff}$, and luminosity, $L$, at $t= 5$ Gyr elapsed from
the zero-age main sequence
from stellar evolution calculations by \citet{yi01,yi03}.
This choice of $t= 5$ Gyr does not affect our calculations of stellar winds,
provided that a star is in the main sequence phase, 
%except for stars with $M_{\star}=0.8M_{\odot}$ and $Z\le 0.01 Z_{\odot}$ 
%the duration of the main sequence is longer than the age of the universe
%\citep[$=13.8$ Gyr;][]{pla16} and
because the stellar properties do not change so much with time.
However, we should note that the duration of
the main sequence of a star with $M_{\star}=0.8M_{\odot}$ and $Z\le 0.01 Z_{\odot}$,
which is $\approx 12-13$ Gyr \citep{mar01}, is slightly shorter than the age
of the universe; if such a star was born at $\approx$
0.1--1 Gyr after the Big Bang, it is currently during the red giant phase. 

The basic stellar parameters, $R_{\star}$, $T_{\rm eff}$, and $L$, of the
zero-metal ($Z=0$) stars in Table \ref{tab:stars_in} were
adopted from the stellar evolution calculations with
$Z=10^{-5}$($\approx 7\times 10^{-4}Z_{\odot}$),
because the effect of different metallicities is quite small for stars
with $Z<0.01 Z_{\odot}$ \citep{sud10}.  
We adopted the radiation cooing for the zero-metallicity gas in the atmosphere
for our stellar wind calculations (\S \ref{sec:cooling}).

The 2nd and 3rd columns of Table \ref{tab:stars_in} show that $T_{\rm eff}$ and
$L$ of lower-metallicity stars are higher for the same stellar
mass. This is because the nuclear fusion energy from the core is effectively
transported by radiation owing to the lower opacity. The velocity amplitude
at the photosphere, which we model later in \S \ref{sec:dv}, is controlled
by $T_{\rm eff}$, and therefore the injected energy depends on stellar
metallicity.

%\subsection{Inner Boundary}

%We inject velocity perturbations from the inner boundary and treat
%the transfer of mass,
%energy, and momentum by solving the propagation of MHD waves in a magnetic
%flux tube. The physical condition at the inner boundary is scaled by a
%reference model for the solar wind presented in our previous works
%\citep{si05,si06,suz13}. 

\subsubsection{Density at Inner Boundary}
The inner boundary of our simulations was set at the location with
$T=T_{\rm eff}$, which is the same as for the solar model.
%The density of the reference case for the solar wind is
%$\rho_{\rm ph,\odot}=10^{-7}$g cm$^{-3}$. We would like to note that, according to
%a model atmosphere calculated by ATLAS \citep{kur79,ck03}, the location
%at $\rho=\rho_{\rm ph,\odot}$ is $\tau_{\rm Ross}\approx 0.1-0.2$, which is
%slightly above the usually defined photosphere at $\tau_{\rm Ross}=2/3$,
%where $\tau_{\rm Ross}$ is the optical depth measured by Rosseland-mean opacity.
We determined the density, $\rho_{\rm ph}$, at the inner boundary by interpolating
ATLAS model atmospheres \citep{kur79,ck03} with different $T_{\rm eff}$,
$Z$, and surface gravity, $g=GM_{\star}/R_{\star}^2$.
%, where $G$ is the gravitational constant.
We note that, since the ATLAS calculations adopted $Z=0.017$ for their
solar abundance from \citet{gs98}, the interpolation is necessary
for $Z$ to fit to the revised value, $Z_{\odot}=0.014$ \citep{asp09}. 
%. In order to fit to the revised value, $Z_{\odot}=0.014$,
%\citep{asp09} we adopt in this paper, we read ATLAS atmosphere data with
%[Fe/H]=0 for $Z=0.017$ when we interpolate data with different metallicities. 
$\rho_{\rm ph}$ of different models are summarized
in Table \ref{tab:stars_in}.
%The density at the photosphere depends on stellar parameters.
%According to calculations on stellar photospheres \citep[\S 9 of][]{gra92},
%the gas pressure at the photosphere has a negative dependence on $T_{\rm eff}$
%and a positive dependence on the surface gravity, $g=GM_{\star}/R_{\star}^2$.
%where G is the gravitational constant and $R_{\star}$ is the stellar radius. 
%From \citet[Fig.9.21 of][\citealt{suz07}]{gra92}, the density at the
%photosphere can be scaled by

We will derive a scaling relation of $\dot{M}$ later in \S \ref{sec:scldotM}.
For this purpose, we would like to present dependence of $\rho_{\rm ph}$
on stellar parameters. 
$\rho_{\rm ph}$ has positive dependences on $g$ and $Z$ and a negative dependence
on $T_{\rm eff}$ \citep[\S 9 of][]{gra92}. The dependences can be roughly fitted
by
\begin{equation}
  \rho_{\rm ph} \propto g^{a}T_{\rm eff}^{-b}D(Z), 
  \label{eq:rhoph}
\end{equation}
with $a=0.55-0.7$ and $b=2-3$ for stars with $4000 \lesssim T_{\rm eff}
\lesssim 6000$ K, where $D(Z)$ is the dependence on metallicity, 
\begin{equation}
D(Z) = 1 + c(1-(Z/Z_{\odot})^{d}) 
\label{eq:Zdeprhoph}
\end{equation}
with $c=2-3$ and $d=0.25-0.3$.
We note that $a$ and $b$ also weakly depend on $Z$; 
numerical fitting of $\rho_{\rm ph}$ from the ATLAS table gives
$a\approx 2$ and $b\approx 0.57$ for $Z=Z_{\odot}$ and $a\approx 8/3$
and $b\approx 0.7$ for $Z\Rightarrow 0$. 
%$D(Z)$ is derived from a fitting to model atmospheres of ATLAS
%\citep[][See Appendix]{kur79,ck03} and $D(0)=4D(Z_{\odot})$. 
%We determine $\rho_{\rm ph}$ at the inner boundary of different stars by this
%scaling from the solar value $\rho_{\rm ph,0}=10^{-7}$g cm$^{-3}$
%\citep{suz07,suz13}.

\subsubsection{Magnetic Field}
We assumed that open magnetic flux tubes on lower-metallicity stars have similar
properties to those on the Sun. 
At the footpoint on the photosphere, we assumed that the magnetic energy is
comparable to the gas energy and the field strength is determined
by equation (\ref{eq:eqpph}).  
In the atmosphere, the flux tube expands, following equation (\ref{eq:ff}).
Here we considered that the loop height, $h_{\rm l}$, is proportional to the
pressure scale height, $H_p$, as a reasonable assumption, and therefore, 
\begin{equation}
  h_{\rm l}\propto H_p \approx c_{\rm s,eff}^2/g \propto T_{\rm eff}/g, 
  \label{eq:psh}
\end{equation}
where $c_{\rm s,eff}=\sqrt{k_{\rm B}T_{\rm eff}/\mu m_{\rm u}}$ is isothermal
sound speed for $T=T_{\rm eff}$. 
$\zeta$ in equation (\ref{eq:ff}) was assumed to
be $\zeta=(1/2)h_{\rm l}$, as was used for the solar case. 

We also adopted the same assumption as the solar case, equation (\ref{eq:avB}),
for the filling factor.
Namely, we assumed the same average magnetic flux density,
$B_{r,0}f_0=1.25$ G, of open magnetic field regions for all the simulated
stars. In other words, these stars have the same magnetic activity level
to the Sun (but see \S \ref{sec:magac} for observed star-to-star variations
of magnetic activity). 
$B_{r,0}$, $h_{\rm l}$, and $f_0$ are tabulated in Table \ref{tab:stars_in}.

\subsubsection{Velocity Perturbation}
\label{sec:dv}
We estimated the amplitude of velocity fluctuations, $\delta v_0$, at the
photosphere from the surface convective flux, which is proportional to
the stellar luminosity $\propto T_{\rm eff}^4$ \citep{ste67,cg68,stp88}: 
\begin{equation}
  \rho_{\rm ph} \delta v_0^3 = \frac{\alpha (\gamma -1)}{2\gamma} \sigma
  T_{\rm eff}^4 \propto T_{\rm eff}^4, 
  \label{eq:dv0}
\end{equation}
where $\alpha$ is the mixing length normalized by the pressure scale height,
which is an order of unity, and $\sigma$ is the Stefan-Boltzmann constant.
%$\gamma$ is the ratio of specific heats.
Equation (\ref{eq:dv0}) determines the scaling relation of $\delta v_0$
of different stars for the reference value, $\langle\delta v_{0,\odot}\rangle
=1.25$ km s$^{-1}$, adopted for the solar case (see \S\ref{sec:dvsun})\footnote{Equation (\ref{eq:dv0}) gives $\delta v_{0,\odot}\approx 4$ km s$^{-1}$ for
  $\rho_{\rm ph}=\rho_{\rm ph,\odot}$, $T_{\rm eff}=T_{\rm eff,\odot}$,
  $\alpha=1.5$, and $\gamma=5/3$. 
  However, we take the smaller value ($=1.25$ km s$^{-1}$) obtained from
  the observation because the photosphere is located slightly above the
  convectively unstable region. 
}.
%{sec:ssm}).
%We adopt our previous simulation for the Sun that well explain the overall
%trend of the current solar wind as the reference, and set $\delta v_0
%=1.34$ km s$^{-1}$ \citep[see][for the detail]{suz13} for a star with
%$M_{\star}=M_{\odot}$, $R_{\star}=R_{\odot}$, and $Z=Z_{\odot}$.
We tabulated the standard value of $\delta v_0$ derived from Equation (\ref{eq:dv0}) in
Table \ref{tab:stars_in}.

We adopted the same spectral shape $\propto \omega^{-1}$ as equation (\ref{eq:dv0})
between $\omega_{\rm min}$ and $\omega_{\rm max}$ with
$\omega_{\rm max}=100\omega_{\rm min}$. 
$\omega_{\rm min}$ and $\omega_{\rm max}$ were scaled by the turnover time of
a typical ``eddy'' ($\approx$ granulation) at the photosphere,
\begin{equation}
  \omega_{\rm min,max}^{-1}\propto H_p/c_{\rm s,eff}\propto c_{\rm s,eff}/g \propto
  \sqrt{T_{\rm eff}}/g 
  \label{eq:tau}
\end{equation}
(see also equation \ref{eq:psh}). 
%where $c_{\rm s}$ is the sound speed and $H_p\approx c_{\rm s}^2/g$
%is the pressure-scale height.
Here we assumed that the typical eddy size
is proportional to $H_p$ at the photosphere. 
The references to the scaling, Equation (\ref{eq:tau}), are $1/\omega_{\rm min} = 30$
min. and $1/\omega_{\rm max}=0.3$ min from the standard solar case
(\S\ref{sec:dvsun}).

\subsubsection{Radiative Cooling}
\label{sec:cooling}
The main coolants in the solar atmosphere are heavy elements. Therefore,
the radiation cooling is suppressed in the atmosphere of lower-mass stars.
We explicitly took into account the metallicity dependence of the radiation
cooling in Equation (\ref{eq:eng}). 
In our code, we combined optically-thin radiation cooling in the coronal region
and optically-thick cooling in the chromosphere; we describe them separately
below. 

\subsubsection*{Corona --Optically-thin cooling}
\label{sec:rcolthin}
\begin{figure}%[h]
  \begin{center}
    \includegraphics[width=0.46\textwidth]{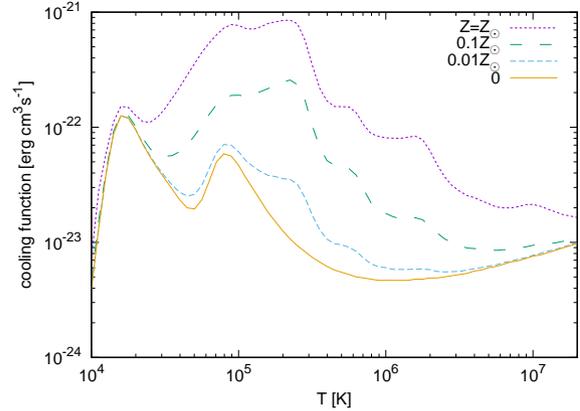}
  \end{center}
  \caption{Cooling functions for optically-thin plasma with different
    metallicities \citep{sd93}. 
  \label{fig:cooling}}
\end{figure}

We adopted tabulated cooling data of optically-thin plasma by \citet{sd93} for
the coronal region with $\rho\le\rho_{\rm cr}=10^{-16}$ g cm$^{-3}$ and
$T\ge 10^{4}$ K. We describe the reason for this choice of
  $\rho_{\rm cr}$ below at the section for the treatment at the transition
  region.
Cooling functions, $\Lambda$ erg cm$^3$s$^{-1}$, for different metallicities
are available (Figure \ref{fig:cooling}).
Volumetric cooling rate $q_{\rm R}$ erg cm$^{-3}$s$^{-1}$ in Equation (\ref{eq:eng}) is calculated via
\begin{equation}
  q_{\rm R}=\Lambda n n_{\rm e},
  \label{eq:radcoolthin}
\end{equation}
where $n$ is ion number density and $n_{\rm e}$ is electron number
density. We assumed fully ionized plasma with mean molecular weight, $\mu=0.6$,
when deriving $n$ and $n_{\rm e}$ from $\rho$.

\subsubsection*{Chromosphere --Optically-thick cooling}
%The cooling of the solar chromosphere with $T\lesssim 10^4$ K is mainly done by
The main coolants in the solar chromosphere are Mg II and Ca II with
smaller contributions from H$\alpha$ and other metallic lines
\citep{ath76,val81}. These lines are not optically thin, and hence it
is necessary to calculate detailed radiative transfer for the accurate
treatment.
In our original code for the solar wind \citep{si05,si06}, instead of
calculating radiation transfer, we adopted an empirical cooling rate,
$q_{\rm R}= 4.5\times 10^{9}\rho$ erg cm$^{-3}$s$^{-1}$, \citep{aa89} derived
from observed chromospheric radiation.
We extended this treatment to lower-metallicity stars.

In a zero-metallicity star, the chromospheric cooling is done
solely by H$\alpha$ emission. 
The observation of the solar chromosphere shows that $\approx 20$\% of
the total chromospheric radiation is from H$\alpha$ \citep{ath76,la78}. 
Following these arguments, we used a simple formula that describes
metallicity-dependent chromospheric cooling rate, 
\begin{equation}
  q_{\rm R} = 4.5\times 10^9\rho\left(0.2+0.8\frac{Z}{Z_{\odot}}\right)
  {\rm erg\; cm^{-3}s^{-1}},
  \label{eq:radcoolthick}
\end{equation}
for the gas with $T\le 10^4$K and $\rho\ge \rho_{\rm cr}$.
%(\S \ref{sec:rcolthin}).
We note that this simplified fitting formula needs to be calibrated by
observations or radiative transfer calculations in future studies. 

\subsubsection*{Transition Region --Interpolation}
The transition region is located between the cool chromosphere and the hot
corona, and its temperature is between $\approx 10^4$ K and $\approx 10^6$ K and
the density is still higher than $\rho_{\rm cr}=10^{-16}$g cm$^{-3}$.
We calculate radiation cooling in the transition region with
  $T>T_{\rm cr}$ and $\rho>\rho_{\rm cr}$ by interpolating Equations
(\ref{eq:radcoolthin}) \& (\ref{eq:radcoolthick}). 
The main reason why we chose $\rho_{\rm cr}=10^{-16}$g cm$^{-3}$ is
somewhat technical; we can connect the two expressions for the radiative
cooling %, equations (\ref{eq:radcoolthin}) \& (\ref{eq:radcoolthick}),
near the bottom of the transition region almost independent from
$Z$; equations
(\ref{eq:radcoolthin}) \& (\ref{eq:radcoolthick}) give the same
value of $q_{\rm R}$ %for $\rho=\rho_{\rm cr}=10^{-16}$g cm$^{-3}$
%near the bottom of the transition region
at $T=T_{\rm cr}=1.2\times 10^4$K for $Z=Z_{\odot}$ or at
$T=T_{\rm cr}=1.1\times 10^4$K for $Z=0$.
%$T=T_{\rm cr}\gtrsim 10^4$K.
$T_{\rm cr}$ depends only weakly on $Z$, because the main coolant is hydrogen
Ly$\alpha$ in this temperature range.

\subsubsection{Initial Condition}

We used the same MHD code described in \S \ref{sec:MHDc} by replacing the
Sun with the lower-metallicity stars in Table \ref{tab:stars_in}.
We set up a static atmosphere with $T=T_{\rm eff}$; in the lower-altitude region
with $\rho \gtrsim 10^{-10} \rho_{\rm ph}$, we adopted the hydrostatic density
structure,
\begin{equation}
\rho_{\rm hs} = \rho_{\rm ph}\exp\left(
-\frac{GM_{\star}}{c_{\rm s,eff}^2}\left(\frac{1}{R_{\star}}-\frac{1}{r}
\right)\right),
\label{eq:rhohs}
\end{equation}
while in the higher-altitude region, we set up density larger than $\rho_{\rm hs}$
in order to avoid unphysically fast \Alfven speed, which causes a troublesome
short time-step when we update physical variables with time.

The simulations were carried out in dimensionless units. The simulation time
is nondimensionalized via $t_{\rm sim}=R_{\star}/c_{\rm s,eff}$. 
We ran the simulations until $t_{\rm sim}=6$, which corresponds to 3-5 times
the sound crossing time and $\gtrsim 10$ times the \Alfven crossing time of
the simulation region. 

\section{Results}
\label{sec:res}
%\subsection{Dependence on Metallicity}

\subsection{Time Evolution: 0.7$M_{\odot}$ Pop.III Star}
\label{sec:tevol}
\begin{figure}[h]
  \begin{center}
        \includegraphics[width=0.48\textwidth]{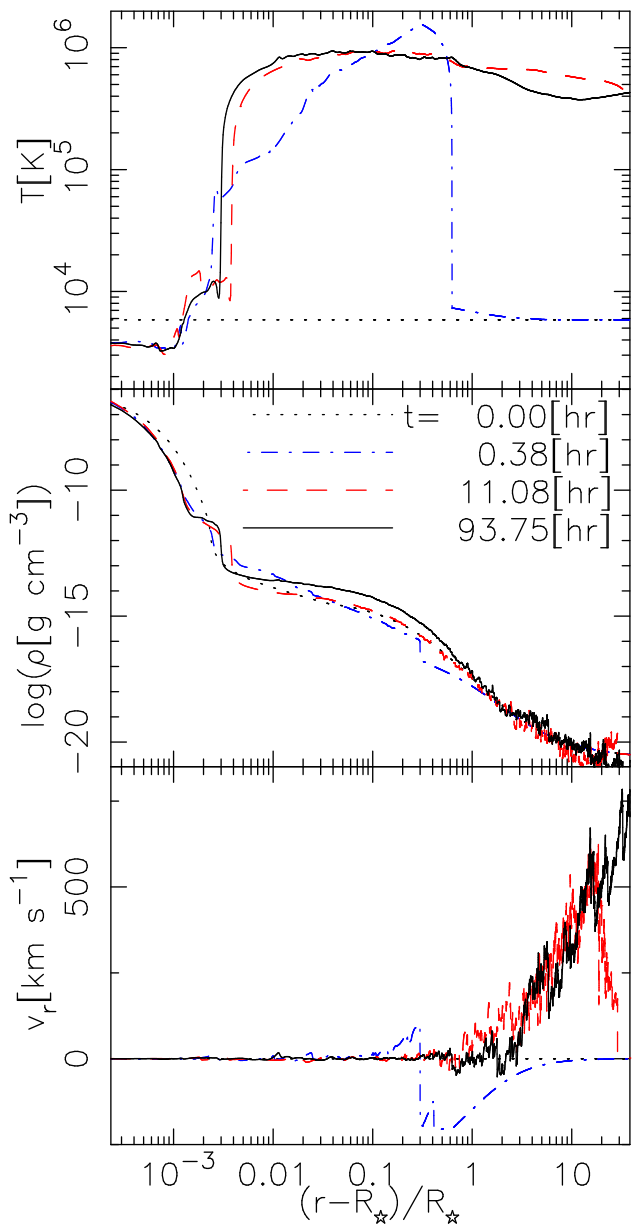}
  \end{center}
  \caption{Time evolution of the atmospheric structure of the star with
    $M_{\star}=0.7M_{\odot}$ and $Z=0$. 
    From the top to the bottom, $T$, $\rho$, and $v_r$ are displayed.
    Black dotted, blue dash-dotted, red dashed and black solid lines
    respectively correspond to the profiles at $t=0$, 0.38. 11.08,
    and 93.75 [hr] from the beginning of the simulation.
    {\it Movie} is also available as a supplementary file and
    at http://ea.c.u-tokyo.ac.jp/astro/Members/stakeru/research/movie/index.html.
  \label{fig:tevol1}}
\end{figure}

Figure \ref{fig:tevol1} demonstrates the time evolution of the atmosphere and
wind of the zero-metal star with $M_{\star}=0.7M_{\odot}$.
The simulation time, $t_{\rm sim}=6$, corresponds to $t\approx100$ hr in the
physical units for this case.

The top panel shows that the upper layer in $r\gtrsim 1.003
R_{\star}$ is quickly heated up to the coronal temperature, $T\approx 10^6$ K,
by the nonlinear dissipation of the \Alfvenic waves from below. 
The main channel of the dissipation is the nonlinear mode conversion; the
fluctuations of the magnetic pressure, $B_{\perp}^2$, with the \Alfvenic waves
excite density perturbations, which propagate as slow-mode MHD ($\approx$
sound) waves. They finally dissipate via shocks that are formed as a result
of the steepening of the wavefronts.
For the detail, see \S \ref{sec:wavedis} and \citet{si05,si06},
\citet{suz13}, and \citet{ms12,ms14}.

The red dash-dotted line at $t=11.08$ hr and black solid line at $t=93.75$ hr
show that the temperature rises with height from $T\approx 4000$ K
to $T\approx 10^4$ K. A nearly isothermal region with
$T\approx 10^4$ K is formed by the Lyman-$\alpha$ cooling, which is seen as
a peak just above $10^4$ K in the cooling curves in Figure \ref{fig:cooling}. 
Therefore, below this quasi-isothermal region, the gas is partially ionized.
%for the partially ionized media).
Above this region, the gas is fully ionized and the temperature jumps up
to $10^6$ K across the sharp transition region because the temperature
range, $T\gtrsim 10^5$ K, is thermally unstable.
Here, we should note that this Ly$\alpha$ plateau may not be realistic
if ambipolar diffusion is properly taken into accuont \citep{fon90}
(see \S \ref{sec:magdif} for the validity of the ideal MHD approximation).

In the middle panel of Figure \ref{fig:tevol1}, the transition region is
recognized as a sharp drop of the density to keep the pressure balance
across this thin layer.
In the chromosphere, the density slightly decreases from the initial value
because the temperature decreases from the initial condition, $T=T_{\rm eff}$.
In other words, the density decreases more rapidly with height because the
pressure scale height ($\propto T$) is shorten.
In contrast, the density in the corona gradually
increases with time by chromospheric evaporation; the chromospheric material is
heated by the downward thermal conduction from the corona and is supplied to
the upper layer. 

The bottom panel of Figure \ref{fig:tevol1} shows that at the beginning
the gas in the upper layer falls down to the surface (blue dash-dotted
lines at $t=0.38$ hr),
since the initial density is larger than $\rho_{\rm hs}$ (equation \ref{eq:rhohs}).
%just sfter we started a simulation at time $t=0$
%with injecting the velocity perturbation (\S \ref{sec:dv}),
However, \Alfvenic waves
propagating from the photosphere push back the infalling gas upward, and
eventually stellar wind streams out in a quasi-steady manner.
%(red dashed line at $t=11.08$ hr and black solid line at $t=93.75$ hr).
After $\approx$ 50 hr, which corresponds to $\approx$ twice the sound crossing
time across the simulation region
$R_{\star}<r\lesssim r_{\rm out}(= 30R_{\star})$, the time-steady velocity
profile is achieved. % in the atmosphere and the wind.

$\rho$ and $v_r$ in the coronal region show fluctuations, most of which are
longitudinal slow MHD (acoustic) wave-like perturbations excited by nonlinear
mode conversion from \Alfvenic waves \citep{ks99}, discussed above.
In contrast, $T$ does not exhibit fluctuations because the thermal conduction
smooths out such small-scale perturbations. 

\begin{figure}%[h]
  \begin{center}
    \includegraphics[width=0.46\textwidth]{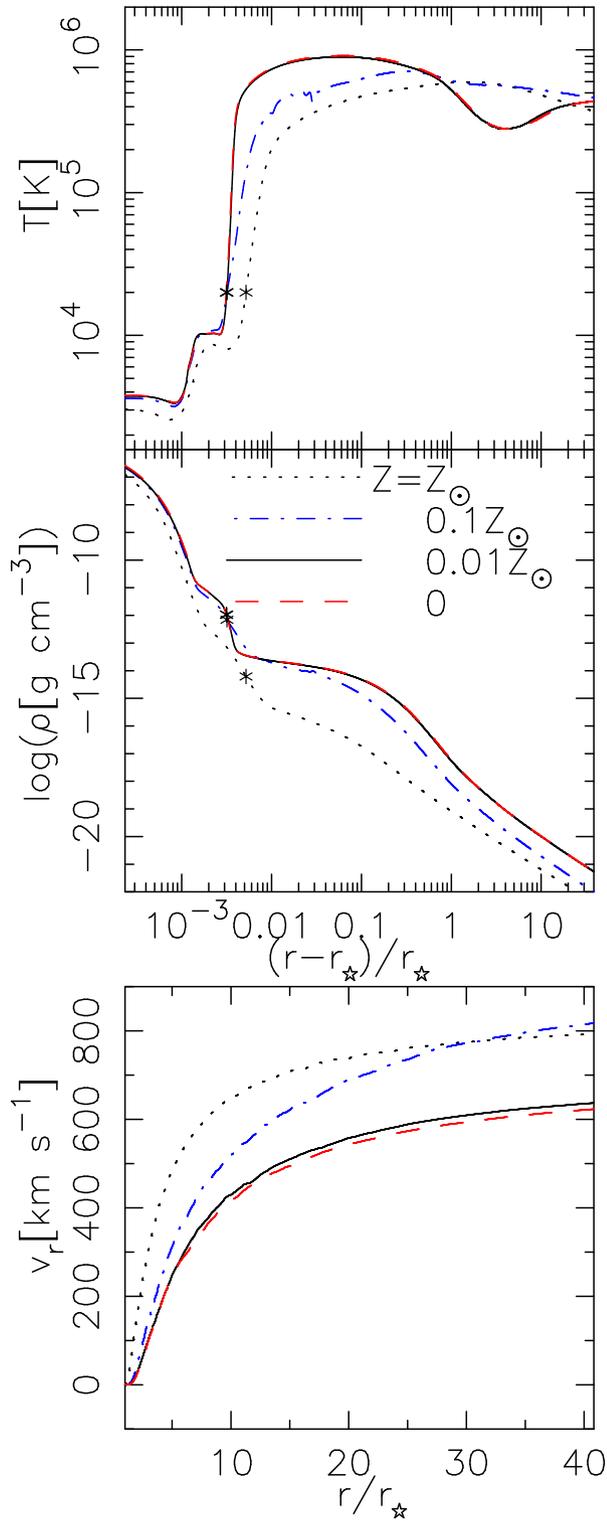}
  \end{center}
  \caption{Comparison of the time-averaged stellar atmosphere and wind
    structures from 0.7 $M_{\odot}$ stars with different metallicities,
    $Z=Z_{\odot}$ (black dotted), $0.1Z_{\odot}$ (blue dash-dotted),
    $0.01Z_{\odot}$ (black solid), and $0$ (red dashed).
    From top to bottom, $T$, $\rho$, and $v_r$ are
    presented. 
    In the top ($T$) and middle ($\rho$) panels, distance from the
    photosphere, $(r-R_{\star})/R_{\star}$ in the logarithmic scale is used to
    zoom in the low-atmospheric region.
    In the top panel ($v_r$), distance from the stellar center, 
    $r/R_{\star}$, is shown in the linear scale for the horizontal axis.
    The asterisk in the top and middle panels indicate the location
    where $T=2\times 10^4$ K. 
  \label{fig:compstr1}}
\end{figure}

\subsection{Dependence on Metallicity}

\begin{figure}%[h]
  \begin{center}
    \includegraphics[width=0.46\textwidth]{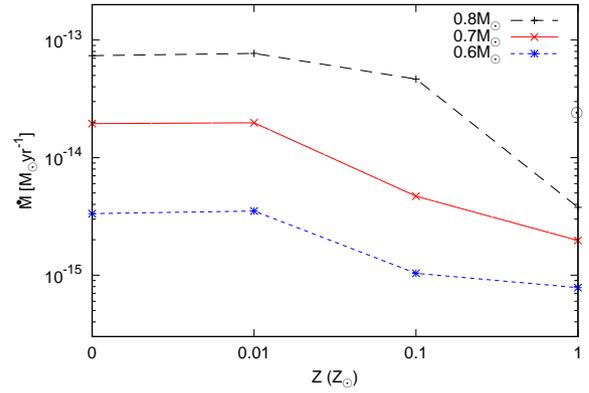}
  \end{center}
  \caption{Dependence of mass loss rates by stellar winds from $0.8 M_{\odot}$
    (black dashed), $0.7 M_{\odot}$ (red solid), and  $0.6 M_{\odot}$ (blue dotted)
    stars on metallicity.
    %The horizontal black dotted line shows the mass loss rate by the solar wind.
    Note that the mass loss rate by the solar wind is also plotted on $Z=Z_{\odot}$. 
  \label{fig:Z-Mdot1}}
\end{figure}

\begin{figure}%[h]
  \begin{center}
    \includegraphics[width=0.46\textwidth]{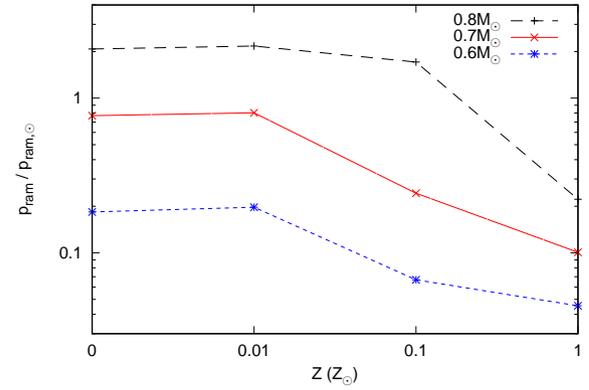}
  \end{center}
  \caption{Dependence of ram pressure of stellar winds at $r=1$ au
    from $0.8 M_{\odot}$ (black dashed), $0.7 M_{\odot}$ (red solid),
    and $0.6 M_{\odot}$ (blue dotted) stars on metallicity.
    $p_{\rm ram}$ is normalized by $p_{\rm ram}$ of the solar case. 
  \label{fig:Z-Pram1}}
\end{figure}

\begin{figure}%[h]
  \begin{center}
    \includegraphics[width=0.46\textwidth]{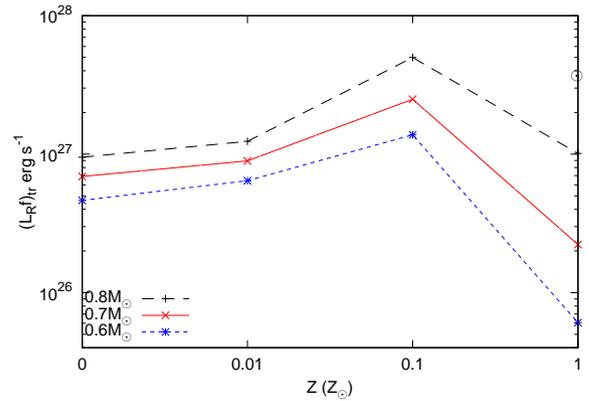}
  \end{center}
  \caption{Dependence of the integrated radiation loss from the open
    magnetic regions in the UV and soft X-ray range, $(L_{\rm R}f)_{\rm tr}$
    (equation \ref{eq:LRf}), of $0.8 M_{\odot}$
    (black dashed), $0.7 M_{\odot}$ (red solid), and  $0.6 M_{\odot}$ (blue dotted)
    stars on metallicity. %The horizontal black dotted line indicates
    $(L_{\rm R}f)_{\rm tr}$ of the solar case is also plotted on
    $Z=Z_{\odot}$. 
  \label{fig:Z-LR1}}
\end{figure}

We investigate how the atmospheres and winds depend on metallicity in this
subsection. Figure \ref{fig:compstr1} compares the atmospheric structures
of 0.7$M_{\odot}$ stars with different metallicities.
Here we focus on the time-averaged structures and took the average
from $t_{\rm sim}=3$ to 6 after the quasi-steady-state structure is achieved. 
The $Z=0.01Z_{\odot}$ case shows the similar profiles of $T$, $\rho$, and $v_r$
to the zero-metallicity case in the three panels, which indicates that
the effect of the different metallicities is negligible on the stellar
winds for $Z\lesssim 0.01Z_{\odot}$.
%on metallicity is almost negligible for $Z\lesssim 0.01Z_{\odot}$. 

The top panel of Figure \ref{fig:compstr1} shows that hot coronae with
temperature $\approx (0.5-1)\times 10^6$ K form in $r\gtrsim (1.005-1.01)
R_{\star}$ in all the four cases.
The temperature profiles are qualitatively similar each other:
A nearly isothermal region with $T\approx 10^{4}$ K is formed by the Ly$\alpha$
cooling, and above that the temperature rapidly rises
owing to the thermal instability (see \S \ref{sec:tevol}). 
However, the peak temperature, $T_{\rm max}$, and its location depend
on metallicity; lower $Z$ gives higher $T_{\rm max}$ that is located closer to
the surface. This is because the efficiency of the cooling is suppressed
for lower $Z$ (Figure \ref{fig:cooling}) and denser gas located at lower
altitudes can be heated up to higher temperature.

%The trasition region is recognized as a sharp drop of the density (the middle
%panel of Figure \ref{fig:compstr1}) to keep the pressure balance across the
%transition region.
The middle panel of Figure \ref{fig:compstr1} indicates that the density
in the coronal region is higher for lower $Z$.
This can be again explained by the suppression of the cooling.
%for smaller $Z$.
As a result, denser gas can be heated up
to the coronal temperature (see also discussion on the energetics later in
\S \ref{sec:energetics}).

The bottom panel of Figure \ref{fig:compstr1} shows that the dependence of
the wind velocity on metallicity is weak, and the terminal velocity
is roughly comparable to the escape velocity, $=\sqrt{2GM_{\star}/R_{\star}}$
($\approx 650$ km s$^{-1}$ for these stars), whereas the wind velocity
is slightly slower for lower $Z$ because denser material has to be lifted up
and accelerated. 

%We compare time-averaged structures of atmospheres and winds from stars with
%different metallicities and masses. The time average is taken during
%$t_{\rm sim}=3$ and 6 in all the cases.
In the last two columns of Table \ref{tab:stars_in}, we show the time averaged
mass loss rate,
\begin{equation}
  \dot{M} = 4\pi r^2\rho v_r f(r),
  \label{eq:mlr}
\end{equation}
and terminal velocity, $v_{\rm t} = v_r$, at $r_{\rm out}(= 30 R_{\star})$,
where $f(r_{\rm out})=1$ in equation (\ref{eq:mlr}). 
The difference of $\dot{M}$ for different $Z$ is 
mostly due to the difference of $\rho$ in the wind region. This is
further connected to the difference of the density at the transition region
marked by asterisks in the middle panel of Figure \ref{fig:compstr1},
where we define this location as the transition region, $r=r_{\rm tr}$, at
$T=2\times 10^4$ K. 
For $M_{\star}=0.7M_{\odot}$, $\dot{M}$ of the $Z\le 0.01Z_{\odot}$ stars
is $\approx$ one order
of magnitude larger than $\dot{M}$ of the solar metallicity star.
This trend is qualitatively similar for different stellar masses,
$0.6-0.8M_{\odot}$, as shown in Figure \ref{fig:Z-Mdot1}. 

Figure \ref{fig:Z-Pram1} compares the ram pressures of stellar winds,
\begin{equation}
  p_{\rm ram} = \rho v_r^2,
\end{equation}
of different $M_{\star}$ cases as a function of $Z$.
Since $p_{\rm ram}$ decreases with $r$, we evaluated it at $r=1$ au, where
we extrapolated $\rho$ and $v_r$ from $r=r_{\rm out}$ to $r=1$ au by assuming
$\rho\propto r^{-2}$ for constant $v_r$.
The vertical axis of Figure \ref{fig:Z-Pram1} is normalized by the value
adopted from our solar case, $p_{\rm ram,\odot}$. Because $p_{\rm ram}\propto
\dot{M} v_{\rm t}$ and $v_{\rm t}$ depends only weakly on $Z$, the general
trend is very similar to that obtained for $\dot{M}$ (Figure \ref{fig:Z-Mdot1}).
The ram pressure is an important parameter to determine the metal pollution
on the surface of low-mass Pop.III stars
\citep[][see also \S \ref{sec:intro}]{tan17}. Figure \ref{fig:Z-Pram1} shows that
$p_{\rm ram}$ for the zero-metal stars with $M_{\star}\ge 0.7M_{\odot}$ is
at least comparable to $p_{\rm ram}$ of the solar wind, and therefore,
the surface pollution is negligible for these stars.

Figure \ref{fig:Z-LR1} compares the integrated radiative coolings in the UV and
soft X-ray range,
\begin{equation}
  (L_{\rm R}f)_{\rm tr} = 4\pi \int_{r_{\rm tr}}^{r_{\rm out}} q_{\rm R}f r^2 dr, 
  \label{eq:LRf}
\end{equation}
of different $M_{\star}$ and $Z$ cases, 
where $r=r_{\rm tr}$ corresponds to the asterisks in Figure \ref{fig:compstr1}. 
%the subscript ``tr'' stands for the transition region at $r=r_{\rm tr}$
%and  $r_{\rm tr}$ is defined as the location where $T=2\times 10^4$K, which corresponds
%to the asterisks in Figure \ref{fig:compstr1}.
The density in the coronal region of the $Z\le 0.01Z_{\odot}$ cases is 1--2 orders
of magnitude larger than that of the solar-metallicity case. The radiative flux
in the corona is proportional to $\rho^2$ (see equation \ref{eq:radcoolthin}).
Although the cooling efficiency, $\Lambda$ erg cm$^3$s$^{-1}$, itself is much smaller
%by 1-1.5 orders of magnitude at $T\approx 10^6$K
for lower $Z$ (Figure \ref{fig:cooling}),
this is totally compensated by the enhanced density.
%We expect that the soft X-ray flux
As a result, $(L_{\rm R}f)_{\rm tr}$ of the $Z\le 0.01Z_{\odot}$ cases is comparable
($M_{\star}=0.8M_{\odot}$) to or even considerably larger ($M_{\star}=0.7M_{\odot}$
and $0.6M_{\odot}$) than that of the solar metallicity case.
$Z=0.1Z_{\odot}$ gives the maximum $(L_{\rm R}f)_{\rm tr}$ for each $M_{\star}$
case, because, compared to the cases with $Z\le 0.01 Z_{\odot}$,
$\Lambda$ is considerably larger although the density is only slightly lower.
We should cautiously note that equation (\ref{eq:LRf}) calculates
the radiative flux from the open magnetic field region. In order to
compare simulated radiative flux to observed UV and X-ray flux, we also
need to take into account the contribution from closed loops, which is
regarded to dominate that from open field regions.

\subsection{Energetics}
\label{sec:energetics}

\begin{figure}%[h]
  \begin{center}
        \includegraphics[width=0.46\textwidth]{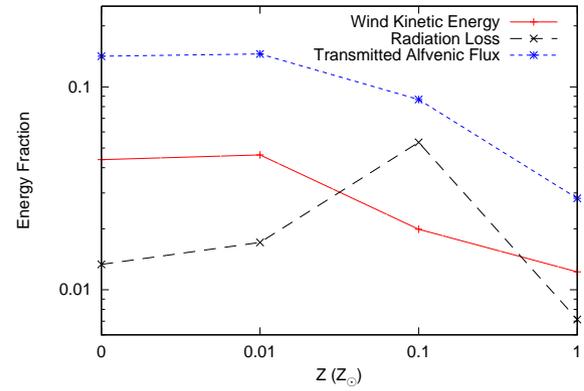}
  \end{center}
  \caption{Comparison of fractional energy flux of the transmitted \Alfvenic
    flux to the transition region (blue dotted), the radiation loss (black
    dotted), and the kinetic energy flux of the stellar wind (red solid)
    of the $0.7M_{\odot}$ stars, normalized by the input \Alfvenic Poynting
    flux from the photosphere as a function of stellar metallicity. 
  \label{fig:Z-Lfrac1}}
\end{figure}

We pursue the metallicity dependence from a more quantitative manner.
In order to do so, we investigate the energetics of the stellar winds.
After various modes of upgoing waves were injected from the photosphere,
the only \Alfvenic (transverse) waves survive into the coronal region,
because compressive
(longitudinal) waves rapidly dissipate by the formation of shocks as a result
of the steepening of the wave fronts \citep[e.g., ][]{suz02}.
Therefore, we focus on the variation of the \Alfvenic Poynting flux with $r$,
and study how the Poynting flux is converted to other types of energy fluxes.

The energy flux of \Alfvenic waves along the $r$ direction is
\citep{jac77,suz13}
\begin{equation}
  F_{\rm A} = v_r\left(\rho\frac{v_{\perp}^2}{2} + \frac{B_{\perp}^2}{4\pi}\right)
  - B_r\frac{v_{\perp}B_{\perp}}{4\pi}, 
\end{equation}
where the first term denotes the energy flux advected by background flow
and the second term indicates the Poynting flux concerning magnetic tension.
The second term dominates the first term in the low atmosphere where the
average flow speed is much smaller than the \Alfven speed, $v_{\rm A}$, while
the opposite is true in the wind region with $v_r > v_{\rm A}$.
Therefore, the injected energy at the photosphere can be expressed as
\begin{equation}
  F_{\rm A,0} \approx - \left(B_r\frac{v_{\perp}B_{\perp}}{4\pi}\right)_0
  \approx \rho_{\rm ph} \langle\delta v_0^2\rangle v_{\rm A,0},
  \label{eq:F_A0}
\end{equation}
where $\langle\cdots\rangle$ stands for the time-average. 
In order to exclude the effect of the adiabatic loss in a super-radially open
flux tube, we introduce \Alfvenic luminosity,
\begin{equation}
  L_{\rm A}f = 4\pi r^2 f F_{\rm A} = \dot{M}\left(\frac{v_{\perp}^2}{2}
  + \frac{B_{\perp}^2}{4\pi\rho}\right) - \Phi_{B}\frac{v_{\perp}B_{\perp}}{4\pi},
  \label{eq:Alflum}
\end{equation}
where $\Phi_B = 4\pi r^2 f B_r$ is the total magnetic flux.

The \Alfvenic waves that travel in the photosphere and chromosphere suffer
reflection because the wave shape is deformed owing to the variation of
$v_{\rm A}$ \citep{wen78,hei80,an90,si06,sho16}, and a small fraction of the input Poynting
flux reaches the corona. We define the transmitted fraction
$=(L_{\rm A}f)_{\rm tr}/(L_{\rm A}f)_0$, where the numerator is evaluated
at $r=r_{\rm tr}$ and the denominator is evaluated at the photosphere, $r=R_{\star}$.
A fraction of the transmitted \Alfvenic energy flux is finally converted
to the kinetic energy flux of the stellar wind,
\begin{equation}
  L_{\rm K,out} = \dot{M}\frac{v_{{\rm t}}^2}{2}. 
  \label{eq:LK}
\end{equation}
%whereas the transmitted Poynting flux is also partly lost via the radiation
%cooling, equation (\ref{eq:LRf}), 
%%where $r_{\rm out}\approx 30 R_{\star}$ is the 
%and is partly used to make the gas escape from the gravitational potential
%\citep{suz13}.
%, where the rest of them travels outward and escapes out of the
%simulation region without converting to other types of energy flux.
%However,

Figure \ref{fig:Z-Lfrac1} shows the fraction of the transmitted \Alfvenic
energy flux, $(L_{\rm A}f)_{\rm tr}/(L_{\rm A}f)_0$, (blue dotted), the radiation
loss at and above the transition region, $(L_{\rm R}f)_{\rm tr}/(L_{\rm A}f)_0$,
(black dashed) and the final kinetic energy flux of the stellar wind,
$L_{\rm K,out}/(L_{\rm A}f)_0$ (red solid) of the $0.7M_{\odot}$ stars. 
We note that the transmitted \Alfvenic energy flux equals to the sum of
the radiation loss (equation \ref{eq:LRf}), the kinetic energy flux (equation
\ref{eq:LK}), the gravitational loss, and the \Alfvenic energy flux outgoing
from $r=r_{\rm out}$ \citep{suz13}, whereas the latter two are not shown. 

The transmitted fraction is $\approx 14\%$ for the low-metallicity
($Z=0.01Z_{\odot}$ and 0) stars and it is $\approx 3\%$ for the solar
metallicity star, which indicates that $\approx 86\%$ or $\approx 97\%$
of the input \Alfvenic energy flux is reflected back downward to the
photosphere in these cases. The transmitted fraction is larger for
lower-metallicity stars because of the difference of the location of
the transition region.
In lower-metallicity stars, dense gas can be heated up to the coronal
temperature because of the suppressed cooling, and therefore, the density at the
transition region is higher. 
As a result, the \Alfvenic waves travel a shorter distance
with a smaller density contrast from the photosphere to the transition region
(the middle panel of Figure \ref{fig:compstr1}), and they
suffer less reflection through the propagation in the chromosphere.

This is, in a sense, a {\it positive feedback} with respect to the heating
by the dissipation of \Alfvenic waves. When metallicity decreases, denser
gas can be heated up by the suppressed radiation cooling, which further
reduces the reflection of \Alfvenic waves.
This raises the transmitted \Alfvenic Poynting flux to the corona,
which further enhances the heating in the corona. 

Since the transmitted fraction is quite small ($\approx 3\%$) in the solar
metallicity star, the fraction of $L_{\rm K,out}$ is also small, $\approx 1.2\%$.
The fraction of $(L_{\rm R}f)_{\rm tr}$ is also tiny; 
the radiation cooling, which is $\propto \rho^2$, is not substantial
because the density at the transition region (and corona) is already
much smaller than in the cases with lower $Z$. 

If we compare the case with $Z=0.1Z_{\odot}$ to the cases with
$Z\le 0.01Z_{\odot}$, a larger fraction is converted to $(L_{\rm R}f)_{\rm tr}$ than
to $L_{\rm K,out}$, because $\Lambda$ in $10^5$K $\lesssim
T \lesssim 10^6$K is larger by a factor of 5-10 (Figure \ref{fig:cooling}).
As a result, the fraction converted to the wind is $L_{\rm K,out}\approx 2\%$,
which is considerably smaller than the fraction $\approx 4.5\%$ obtained in the
lower $Z$ cases.

%The kinetic energy flux is roughly proportional to the mass loss rate
%(equation \ref{eq:LK}), since the wind velocities are not so different in different
%cases.
Table \ref{tab:stars_in} and Figure \ref{fig:Z-Mdot1} show that $\dot{M}$ of
$Z\le 0.01 Z_{\odot}$ is 10 times and $\dot{M}$ of $Z=0.1Z_{\odot}$ is 2.4 times
larger than $\dot{M}$ of $Z=Z_{\odot}$ for $M_{\star}=0.7M_{\odot}$. These values
are larger than the estimates from the energy conversion efficiency,
$L_{\rm K,out}/(L_{\rm A}f)_0$, we have discussed above. 
%In addtion to the larger fraction of $L_{\rm K,out}$,
This is mainly because the input \Alfvenic
energy, $(L_{\rm A}f)_0$, from the photosphere (equation \ref{eq:F_A0}) is larger
for lower $Z$ on account of the moderately larger convective flux
($\propto T_{\rm eff}^4$; see equation \ref{eq:dv0} and Table \ref{tab:stars_in}); 
%$\rho_{\rm eff}$ and $\delta v_0$ are
$(L_{\rm A}f)_0$ of $Z\le 0.01 Z_{\odot}$ is 1.7 times and $(L_{\rm A}f)_0$ of
$Z= 0.1 Z_{\odot}$ is 1.5 times larger than $(L_{\rm A}f)_0$ of $Z=Z_{\odot}$.
In addition, the terminal velocity, $v_{\rm t}$, of the $Z\le 0.01 Z_{\odot}$
cases is $\approx 20\%$ slower than $v_{\rm t}$ of the $Z=Z_{\odot}$ case.
Therefore, the difference ($\approx$ a factor of 10) of $\dot{M}$
between $Z\le 0.01 Z_{\odot}$ and $Z=Z_{\odot}$ is larger than the difference
($\approx$ a factor of 6-7) of $L_{\rm K,out}$ (equation \ref{eq:LK}). 
%Hence, $\dot{M}$ of the $Z\le 0.01Z_{\odot}$ cases is
%$\approx 4$ times larger than $\dot{M}$ of the $Z=0.1Z_{\odot}$ case,
%and $\approx 10$ times larger than $\dot{M}$ of the $Z=Z_{\odot}$ case. 

%\begin{figure}%[h]
%  \begin{center}
%    \includegraphics[width=0.46\textwidth]{../work/Z-Lfrac_07Mdv201.eps}
%  \end{center}
%  \caption{Dependence of mass loss rate on metallicity for $\delta v\times 2$.
%  \label{fig:Z-Mdot1}}
%\end{figure}

%dv x 2の時には、抜けていくPoynting Fluxが減り、ほとんど全てが加熱に使
%われるので$L_{\rm K}$, $L_{\rm R}$の割合が高い。

%\subsection{Dependence on Mass}
%\begin{figure}%[h]
%  \begin{center}
%    \includegraphics[width=0.46\textwidth]{../work/M-Mdot_0Z_01.eps}
%  \end{center}
%  \caption{Dependence of mass loss rate on metallicity.
%  \label{fig:Z-Mdot1}}
%\end{figure}

%\begin{figure}%[h]
%  \begin{center}
%    \includegraphics[width=0.46\textwidth]{../work/M-Lfrac_Z0dv101.eps}
%  \end{center}
%  \caption{Dependence of mass loss rate on metallicity for $\delta v\times 1$.
%  \label{fig:Z-Mdot1}}
%\end{figure}

\subsection{Scaling Relation for $\dot{M}$}
\label{sec:scldotM}

\begin{figure}[h]
  \begin{center}
    \includegraphics[width=0.46\textwidth]{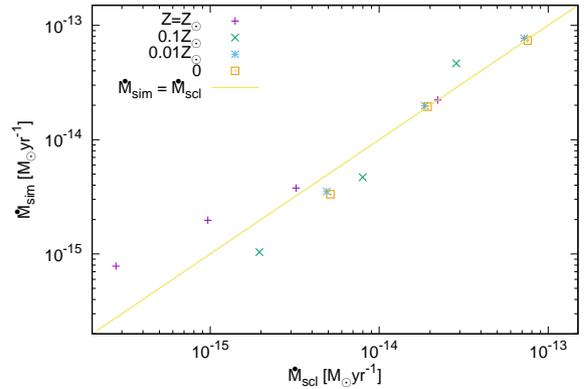}
  \end{center}
  \caption{Comparison of the mass loss rate from the scaling relation,
    $\dot{M}_{\rm scl}$ (equation \ref{eq:dotMscaling2}), and the mass loss rate
    of the simulations, $\dot{M}_{\rm sim}$. The normalization of
    the scaling is from the solar value $\eta_{\rm scl}=2.22\times 10^{-14}
    M_{\odot}$yr$^{-1}$ (Table \ref{tab:stars_in}).
    Different symbols correspond to
    different metallicities. As for the solar metallicity
    cases, we show the result of $M_{\star}=M_{\odot}$ in addition
    to $M_{\star}=0.6$, 0.7, \& 0.8$M_{\odot}$.  
    \label{fig:Mdotscale}}
\end{figure}

In this subsection we derive a simple scaling relation of $\dot{M}$
from our simulations,
following the energetics argument in \S \ref{sec:energetics}.
Figure \ref{fig:Z-Lfrac1} indicates that the wind kinetic energy is almost
proportional to the transmitted \Alfvenic Poynting flux to the corona, namely,
$L_{\rm K,out}\propto (L_{\rm A}f)_{\rm tr}$. The terminal velocity can be roughly
scaled by the escape velocity, $v_{\rm t}\sim \sqrt{2GM_{\star}/R_{\star}}$, and
then, we have
\begin{equation}
  \dot{M}\propto (L_{\rm A}f)_0 c_{\rm T}(R_{\star}/M_{\star}), 
  \label{eq:dotMscaling}
\end{equation}
where $c_{\rm T}=(L_{\rm A}f)_{\rm tr}/(L_{\rm A}f)_0$ is the transimissivity of
\Alfvenic waves from the photosphere to the corona.
Using equations (\ref{eq:dv0}) \& (\ref{eq:F_A0}), we get the dependence of
$(L_{\rm A}f)_0$ on stellar parameters:
\begin{eqnarray}
  (L_{\rm A}f)_0&=&4\pi R_{\star}^2f F_{\rm A,0} \nonumber \\
  &\propto& \rho_{\rm ph}^{1/2}\delta v_0^2
  (B_rf)_0 R_{\star}^2 \nonumber \\
  &=&(\rho_{\rm ph}\delta v_0^3R_{\star}^2)^{2/3}\rho_{\rm ph}^{-1/6}R_{\star}^{2/3}
  \nonumber \\ &\propto& L^{2/3}\rho_{\rm ph}^{-1/6}R_{\star}^{2/3}, 
  \label{eq:LAfscaling}
\end{eqnarray}
where we used the assumption, $(B_rf)_0=$const. in equation (\ref{eq:avB}).

After \Alfvenic waves are excited from the photosphere, these waves, which are
affected by dissipation and reflection, travel upward. Determining $c_{\rm T}$ is
a very difficult task because it is not simple to properly take into
account these processes with nonlinear effects.
Here we consider the situation in which the reflection is a dominant process
that controls $c_{\rm T}$. 
In the stellar atmosphere, both density and magnetic field strength change
rapidly with height. As a result, the \Alfven velocity also varies.
\Alfvenic waves with wavelength, $\lambda$, longer than the variation scale of $v_{\rm A}$
are subject to reflection because of the deformation of the wave shape
\citep{si06,sho16}. In the long-wavelength limit, $\lambda \gg
|\frac{d r}{d\ln v_{\rm A}}|$, we can derive the transmitted wave amplitude from
a region I with density $\rho_{\rm I}$ to a region II with $\rho_{\rm II}$ as
follows \citep{hol84,ver12}:
\begin{equation}
  \delta v_{\perp,{\rm II}} = \frac{2}{1+\sqrt{\rho_{\rm II}/\rho_{\rm I}}}
  \delta v_{\perp,{\rm I}}
  \label{eq:dvtransm}
\end{equation}
Using this transmissivity for velocity amplitudes, we can estimate $c_{\rm T}$
from the photosphere to the corona: 
\begin{equation}
  c_{\rm T}=\frac{(\rho\delta v_{\perp}^2 v_{\rm A,tr}f r^2)_{\rm tr}}
  {\rho_{\rm ph}\delta v_0^2 v_{\rm A,0}f_0 R_{\star}^2}
  = \frac{4\sqrt{\rho_{\rm ph}/\rho_{\rm tr}}}
  {(\sqrt{\rho_{\rm ph}/\rho_{\rm tr}}+1)^2}\approx 4\sqrt{\frac{\rho_{\rm tr}}
    {\rho_{\rm ph}}}, 
\end{equation}
where $\rho_{\rm tr}$ is the density at the bottom of the transition
region at which $T=T_{\rm tr}$($=2\times 10^4$K).
We used $(\sqrt{\rho}v_{\rm A}fr^2)_{\rm tr} = \sqrt{\rho_{\rm ph}}
v_{\rm A,0}f_{0}R_{\star}^2$ derived from the conservation of magnetic flux,
equation (\ref{eq:divB0}), and $\rho_{\rm ph}\gg \rho_{\rm tr}$ for the last
approximate equality.

%The density at the transition region,
$\rho_{\rm tr}$ is determined by the
balance between heating and conductive and radiative cooling \citep{rtv78};
the heating in the transition region and the low corona is mainly lost by
the downward thermal conduction to the chromosphere, in addition to
the radiative cooling.  When heating increases, the enhanced downward thermal
conduction makes cool chromospheric materials evaporate to the corona, which
leads to larger $\rho_{\rm tr}$.
Since the conductive flux is finally lost by the radiation in the transition
region and the upper chromosphere, we can assume that the heating by the wave
dissipation, $(\rho\delta v_{\perp}^2v_{\rm A})_{\rm tr}/\tau_{\rm dis}$, is balanced
by $q_{\rm R}$ in equation (\ref{eq:eng}). We adopt optically thin cooling,
equation (\ref{eq:radcoolthin}), and then, $q_{\rm R}\propto \rho^2 \Lambda$.
Introducing a dissipation length, $l_{\rm dis}$, we can write the heating rate
by the dissipation of \Alfvenic waves
as $\rho\delta v_{\perp}^2v_{\rm A}/l_{\rm dis}$. Then, we obtain 
an equation that describes the energy balance at and above the transition
region, 
\begin{equation}
  \int_{r_{\rm tr}}^{r_{\rm max}}dr\left(\frac{\rho}{\mu m_{\rm u}}\right)^2
  \Lambda \sim \int_{r_{\rm tr}}^{r_{\rm max}}dr
  \frac{\rho\delta v_{\perp}^2v_{\rm A}}{l_{\rm dis}}, 
  \label{eq:engbalance}
\end{equation}
where the integration is done from the bottom of the transition region,
$r=r_{\rm tr}$, to the location, $r=r_{\rm max}$, that gives the maximum
temperature. 
This relation indicates that enhanced heating and/or reduced cooling
leads to higher $\rho_{\rm tr}$ owing to the efficient chromospheric
evaporation, as explained above. 
We can transform the left-hand-side of equation (\ref{eq:engbalance}) as
\begin{equation}
  \int_{r_{\rm tr}}^{r_{\rm max}}dr \rho^2 \Lambda \equiv \langle \Lambda
  \rangle_{\rho^2}\int_{r_{\rm tr}}^{r_{\rm max}}dr \rho^2 \approx \langle \Lambda
  \rangle_{\rho^2}\rho_{\rm tr}^2H_{\rm tr},
  \label{eq:rhoweightLambda}
\end{equation}
where for the last approximate equality we used the fact that the integral
of density is heavily weighted on the smaller $r$ side near $r=r_{\rm tr}$
because the density rapidly decreases with $r$, and $H_{\rm tr}=k_{\rm B}
T_{\rm tr}/\mu m_{\rm u}g$ is the pressure scale height measured
for $T=T_{\rm tr}$($=2\times 10^4$K). We put $\rho^2$ for the subscript of
$\langle \Lambda \rangle$ to explicitly show that this is the
${\rho}^2$-weighted cooling function.

In our simulations, \Alfvenic waves dissipate via nonlinear processes (see
\S\ref{sec:wavedis}). In this case, we can model that the dissipation rate
is proportional to nonlinearity, $\delta v_{\perp}/v_{\rm A}$, and that 
\begin{equation}
  l_{\rm dis}\sim \lambda(\delta v_{\perp}/v_{\rm A})^{-1}
  \sim \frac{v_{\rm A}^2}{\omega \delta v_{\perp}},
  \label{eq:ldis}
\end{equation}
where we used $\lambda\sim v_{\rm A}/\omega$. 
From equation (\ref{eq:dvtransm}), we have
\begin{equation}
  \delta v_{\perp,{\rm tr}}\approx 2\delta v_0 \propto \delta v_0
  \label{eq:dvscale}
\end{equation}
The integration of the right-hand-side of equation
(\ref{eq:engbalance}) is also weighted near $r=r_{\rm tr}$ similarly
to the left-hand side (equation \ref{eq:rhoweightLambda}) because the
volumetric heating rate is proportional to $\rho$.
Substituting equations (\ref{eq:ldis}) \& (\ref{eq:dvscale}) into
the the right-hand-side of equation (\ref{eq:engbalance}), we have
\begin{equation}
  \int_{r_{\rm tr}}^{r_{\rm max}}dr
  \frac{\rho\delta v_{\perp}^2v_{\rm A}}{l_{\rm dis}}
  \approx \frac{\rho_{\rm tr}\delta v_0^3\omega}{v_{\rm A,tr}} H_{\rm tr}
  \label{eq:rhoweightheating}
\end{equation}
Applying equations (\ref{eq:rhoweightLambda}) \& (\ref{eq:rhoweightheating})
to equation (\ref{eq:engbalance}), we obtain
\begin{eqnarray}
  \rho_{\rm tr}^2 \langle \Lambda \rangle_{\rho^2} &\propto& \rho_{\rm tr}
  \delta v_0^3\omega/v_{\rm A,tr}
  \nonumber \\
  &\sim& \rho_{\rm tr}^{3/2}T_{\rm eff}^4\rho_{\rm ph}^{-1}M_{\star}R_{\star}^{-2}
  T_{\rm eff}^{-1/2} \nonumber \\
  &=&\rho_{\rm tr}^{3/2}T_{\rm eff}^{7/2}\rho_{\rm ph}^{-1}M_{\star}R_{\star}^{-2},
\end{eqnarray}
where we used equations (\ref{eq:dv0}) \& (\ref{eq:tau}), and $v_{\rm A,tr}\propto
\rho_{\rm tr}^{-1/2}$ because we can assume $B_{r,{\rm tr}}\sim (Bf)_0=$const.
As a result, we get the scaling of $c_{\rm T}$, 
\begin{equation}
  c_{\rm T}\propto \langle\Lambda\rangle_{\rho^2}^{-1}T_{\rm eff}^{7/2}
  M_{\star}R_{\star}^{-2}\rho_{\rm ph}^{-3/2}.
  \label{eq:ctscaling}
\end{equation}
From equation (\ref{eq:rhoph}), we adopt
\begin{equation}
  \rho_{\rm ph}\propto M_{\star}^{2/3}R_{\star}^{-4/3}T_{\rm eff}^{-2},
  \label{eq:rhophscale}
\end{equation}
where we neglected the weak dependence on metallicity. 
Substituting equations (\ref{eq:LAfscaling}), (\ref{eq:ctscaling}) \&
(\ref{eq:rhophscale}) into equation (\ref{eq:dotMscaling}), we finally have
the scaling of $\dot{M}$:
\begin{eqnarray}
  \dot{M}&\propto&\langle\Lambda\rangle_{\rho^2}^{-1}LR_{\star}^{-1}T_{\rm eff}^{13/6}\rho_{\rm ph}^{-5/3}
  \nonumber \\
  &\propto& \langle\Lambda\rangle_{\rho^2}^{-1}LR_{\star}^{11/9}M_{\star}^{-10/9}T_{\rm eff}^{11/2}
  \label{eq:dotMscaling2}
\end{eqnarray}
The cooling function, $\Lambda$, depends on $Z$ (Figure \ref{fig:cooling}).
If we focus on gas in the transition region with $10^{4}\lesssim T \lesssim 10^6$ K,
$\Lambda$ does not depend on $Z$ for $T\lesssim 10^{4.5}$K because the main
coolant is hydrogen atoms, while $\Lambda$ strongly depends on $Z$ for
$T\gtrsim 10^{4.5}$K. Although the
former temperature range is quite narrow, it is not negligible for the
  $\rho^2$-weighted cooling, $\langle\Lambda\rangle_{\rho^2}$ because it
  %the cooling rate, $q_{\rm R}\propto \rho^2$,
is biased on the lower temperature  
side. Considering this situation, we adopt a weak dependence on $Z$ with
a floor in $Z<0.01Z_{\odot}$, which can reasonably explain our simulation
results:
\begin{equation}
  \langle\Lambda\rangle_{\rho^2}\propto \left[\max\left(\frac{Z}{Z_{\odot}},0.01\right)\right]^{\frac{1}{5}}. 
  \label{eq:coolonZ}
\end{equation}
Applying equation (\ref{eq:coolonZ}) to equation (\ref{eq:dotMscaling2}),
we finally obtain an equation that predicts mass loss rate from stellar basic
parameters:
\begin{eqnarray}
  \dot{M}_{\rm scl} &=& \eta_{\rm scl}\frac{L}{L_{\odot}}  
  \left(\frac{R_{\star}}{R_{\odot}}\right)^{\frac{11}{9}}
  \left(\frac{M_{\star}}{M_{\odot}}\right)^{-\frac{10}{9}}
  \left(\frac{T_{\rm eff}}{T_{\rm eff,\odot}}\right)^{\frac{11}{2}}\nonumber \\
  & &\times \left[\max\left(\frac{Z}{Z_{\odot}},0.01\right)\right]^{-\frac{1}{5}}, 
  \label{eq:dotMscaling3}
\end{eqnarray}
where the normalization is adopted from the solar mass loss rate of
our simulation, $\eta_{\rm scl}=\dot{M}_{\odot}$($=2.22\times 10^{-14}M_{\odot}
$yr$^{-1}$; Table \ref{tab:stars_in}).

It is worth comparing this relation to previous works.     
The famous Reimers' relation \citep{rei75} was derived from simple energetics,
\begin{equation}
  \dot{M}_{\rm Reimers} = \eta_{\rm Reimers}\frac{(L/L_{\odot})
    (R_{\star}/R_{\odot})}{M_{\star}/M_{\odot}}, 
  \label{eq:Reimers}
\end{equation}
where the original normalization, $\eta_{\rm Reimers}=4\times 10^{-13}
M_{\odot}$yr$^{-1}$, was adopted as a standard value for red supergiants.
    
\citet{sc05} modified this relation by including an mechanical energy input and
a chromospheric radius, which is located far above the photosphere in red giant
stars, and derive
\begin{equation}
  \hspace{-0.5cm}\dot{M}_{\rm SC05}= \eta_{\rm SC05} \frac{(L/L_{\odot})(R_{\star}/R_{\odot})}
      {M_{\star}/M_{\odot}}\left(\frac{T_{\rm eff}}{4000\;{\rm K}}\right)^{3.5}
      \left(1+\frac{g_{\odot}}{4300g}\right), 
      \label{eq:SC05}
\end{equation}
where $\eta_{\rm SC05}=8\times 10^{-14}M_{\odot}$yr$^{-1}$ was introduced
as a standard normalization for red giants. 
The power-law indices of $R_{\star}$, $M_{\star}$, and $T_{\rm eff}$ in
equation (\ref{eq:dotMscaling3}) are slightly different from those in
equation (\ref{eq:SC05}).
The biggest difference is that our relation explicitly considers the effect of
metallicity. If we are to apply our relation to metal-poor red giants, it is
probably better to include the explicit dependence on $g$ presented in
equation (\ref{eq:SC05})

Figure \ref{fig:Mdotscale} compares $\dot{M}_{\rm scl}$ derived from
equation (\ref{eq:dotMscaling3}) and $\dot{M}_{\rm sim}$ of the numerical simulations.
Although we took several crude simplifications, the derived scaling relation
explains the overall trend of the simulation results.
The fitting of either higher-mass ($M_{\star}=0.8M_{\odot}$) or lower-metallicity
($Z\le 0.01Z_{\odot}$) cases is quite nice.
On the other hand, $\dot{M}_{\rm scl}$ of lower-mass ($M_{\star} \le 0.7M_{\odot}$)
solar metallicity stars underestimates
$\dot{M}_{\rm sim}$, while $\dot{M}_{\rm scl}$ of $M_{\star}=0.6M_{\odot}$ and
$Z=0.1Z_{\odot}$ slightly overestimates $\dot{M}_{\rm sim}$. 

\section{Discussion}
\label{sec:dis}

\subsection{Magnetic Diffusion}
\label{sec:magdif}
We should critically check the validity of the ideal MHD approximation we
have assumed in this paper, because the gas in the photosphere and chromosphere
is not fully ionized. While in the solar metallicity gas metals with low
ionization potential dominantly supply electrons, in the zero-metal gas the
hydrogen is almost the sole source of electrons in the photosphere and
chromosphere because the helium stays as neutral.
%Therefore, the ionization degree is lower for lower $Z$ and \Alfven waves tend
%to be damped by non-ideal MHD processes.
On the other hand, the effective temperature of a lower metallicity star
is higher than that of a higher metallicity star (Table \ref{tab:stars_in}). 
If we compare two stars with the same $M_{\star}$ but different $Z$, the former
effect decreases the ionization degree of the lower $Z$ star, while
the latter effect increases it. If we take the four stars with $M_{\star}
= 0.7M_{\odot}$ in Table \ref{tab:stars_in} for example, these stars give similar
ionization degrees, $x_e\sim 10^{-5} - 10^{-4}$ at the photosphere
under the local-thermodynamical-equilibrium condition, because the
two effects are almost canceled out each other.

In the high-density condition, Ohmic dissipation is the dominant damping
mechanism. Ohmic resistivity (magnetic diffusivity), $\chi_{\rm O}$, by
electron-neutral collision can be estimated \citep{bb94} as
\begin{equation}
  \chi_{\rm O} \approx 230 \sqrt{T}/x_e \; {\rm cm^2 s^{-1}}.
  \label{eq:etaO}
\end{equation}
%where $x_e$ is the ionization degree.
We evaluate how the Ohmic dissipation affects the propagation of \Alfven
waves by a magnetic Reynolds number. In fluid mechanics, a Reynolds number,
$Re$, which is defined as the ratio of an inertial term to a viscous term, 
is often used as a measure of dissipation; flow tends to be laminar for
$Re\lesssim 1$ and turbulent for $Re\gg 1$. In order to examine the propagation
of \Alfvenic Poynting flux, an inertial term is replaced by a term derived from
\Alfven waves,
\begin{equation}
  C_{\rm A}=\lambda v_{\rm A} = 10^{14}\left(\frac{\lambda}{1000\;{\rm km}}\right)
  \left(\frac{v_{\rm A}}{10\;{\rm km\;^{-1}}}\right)\; {\rm cm^2 s^{-1}}, 
  \label{eq:CA}
\end{equation}
where %$\lambda$ is the wavelength, and
$\lambda(=v_{\rm A}/\omega)$ and $v_{\rm A}$ were normalized by typical quantities
in the photosphere and chromosphere. In the zero-metal star with $M_{\star}
= 0.7M_{\odot}$, we inject perturbations in the frequency range between
$\omega_{\rm max}^{-1}\approx 10$ s and $\omega_{\rm min}^{-1}\approx 1000$ s.
$\lambda=1000$ km corresponds to the central value, $\omega^{-1}=100$ s.

Under the thermal equilibrium, the Saha equation for the ionization of hydrogen
atoms gives $x_e\propto 1/\sqrt{\rho}$ \citep[e.g., ][]{gra92}.
Hence, the Ohmic
diffusion affects most severely at the densest location, namely the photosphere (inner boundary) in our simulations. 
From equations (\ref{eq:etaO}) and (\ref{eq:CA}), we estimate a magnetic Reynolds
number by the Ohmic dissipation at the photosphere of
the zero-metal star with $M_{\star} = 0.7M_{\odot}$,
\begin{eqnarray}
  Re_{\rm O} &=&\frac{C_{\rm A}}{\chi_{\rm O}} \nonumber \\
  &\approx& 4.0\times 10^{5}
  \left(\frac{x_e}{7\times 10^{-5}}\right)\left(\frac{T}{5842\;{\rm K}}
  \right)^{-1/2}\nonumber \\
  & & \hspace{1cm}\left(\frac{\lambda}{1000\; {\rm km}}\right)
  \left(\frac{v_{\rm A}}{10\;{\rm km\;s^{-1}}}\right),   
\end{eqnarray}
where the normalization of $x_e =7\times 10^{-5}$ is adopted from
(the interpolation of) ATLAS atmospheres, and this is consistent with
the value derived from the Saha equation for the ionization of hydrogen atoms.
Although $Re_{\rm O}$ becomes an order of magnitude smaller for
$\omega_{\rm max}^{-1}\approx 10$ s, it still gives $Re_{\rm O}
\approx 4.0\times 10^{4} \gg 1$. Lower-mass stars give lower ionization
because the temperature is lower; the zero-metal star with $0.6M_{\odot}$
gives $x_e=1.2\times 10^{-5}$. $Re_{\rm O}$ estimated from this $x_e$ is
still much larger than unity.
Therefore, we can conclude that the \Alfven waves we have considered are
not so affected by the Ohmic diffusion. 

In the low-density condition, ambipolar diffusion between charged particles
and neutral particles is the main damping
mechanism. Ambipolar diffusivity, $\chi_{\rm A}$, can be estimated from $x_e$
and ion-neutral collision rate \citep{nu86,sus15} as
\begin{equation}
  \chi_{\rm A} = \frac{(m_i + m_n) B^2}{4\pi\langle \sigma v\rangle_{\rm in}\rho_i
    \rho_n},
  \label{eq:etaA}
\end{equation}
where the subscript $i$ and $n$ denote ions and neutrals, respectively, and
$\langle \sigma v\rangle_{\rm in} = 1.9\times 10^{-9}$cm$^3$s$^{-1}$ is ion-neutral
collision rate\footnote{Although we used $\sigma$ for the Stefan-Boltzmann
  constant in equation (\ref{eq:dv0}), we also adopted $\sigma$ for the cross
  section in equation (\ref{eq:etaA}) because it does not cause confusion.}
per number density \citep{dra83}. Substituting
$m_{i}=m_{\rm H}$ and $m_n=\mu m_{\rm H}$ with $\mu=1.3$, we can derive for
$x_e\ll 1$ 
\begin{eqnarray}
  \chi_{\rm A} &\approx& 2.1\times 10^{-16}\frac{B^2}{\rho^2 x_e}\; {\rm cm^2 s^{-1}} \nonumber \\
  &=& 2.1\times 10^{11}\left(\frac{B}{100\;{\rm G}}\right)^2 \nonumber \\
  & &\hspace{1cm} \left(\frac{\rho}{10^{-10}{\rm g\; cm^{-3}}}\right)^{-2}
  \left(\frac{x_e}{10^{-3}}\right)^{-1} \; {\rm cm^2 s^{-1}},
  \label{eq:etaA}
\end{eqnarray}
where the normalizations are adopted from the physical quantities at $T\approx
5000$ K in the mid to upper chromosphere; the ambipolar diffusion is expected
to affect the propagation of \Alfven waves most substantially in this region
because the density becomes low but the ionization is still not high there. 

From equations (\ref{eq:CA}) and (\ref{eq:etaA}), we can estimate a magnetic Reynolds
number by the ambipolar diffusion in the mid to upper chromosphere,
\begin{eqnarray}
  Re_{\rm A} &=&\frac{C_{\rm A}}{\chi_{\rm A}} \nonumber \\
  &\approx& 4.8\times 10^2 \left(\frac{B}{100\;{\rm G}}\right)^2 
  \left(\frac{\rho}{10^{-10}{\rm g\; cm^{-3}}}\right)^{-2}
  \nonumber \\
  & &\hspace{1cm}\left(\frac{x_e}{10^{-3}}\right)^{-1}
  \left(\frac{\lambda}{1000\; {\rm km}}\right)
  \left(\frac{v_{\rm A}}{10\;{\rm km\;s^{-1}}}\right).
\end{eqnarray}
This estimate is again for the \Alfven wave with $\omega^{-1}=100$ s, and for
higher-frequency waves, $\omega_{\rm max}^{-1}=10$ s, $Re_{\rm A}=48$.
The ionization degree is lower for lower-mass stars.
For example, $Re_{\rm A}\approx 10$ for the \Alfven wave with
$\omega=\omega_{\rm max}$ in %the chromosphere of the zero-metal star with
$M_{\star}=0.6M_{\odot}$. Although $Re_{\rm A}$ is still larger
than unity, the \Alfven waves in the higher-frequency range within the injected
spectral band are regarded to be subject to damping.
For more realistic treatment for these stars in our future studies, we need
to take into account ambipolar diffusion in the chromosphere. 

\subsection{Magnetic Activity of Metal-poor Stars}
\label{sec:magac}
A crucial assumption of our work is that we determined the properties of the
magnetic flux tubes by extrapolating the parameters of a typical magnetic
flux tube on the Sun, because we have no direct observational information on
the magnetic field of low-mass Pop. II/III stars. 
Although a Pop.III star has not been discovered, we infer some clues of 
magnetic activity of low-mass Pop.II stars from observations by UV and X-ray
radiation. 

\citet{ott97} analyzed X-ray observations of 86 Pop.II binaries from the ROSAT
all-sky survey. They detected X-rays from the stellar coronae of 13 systems
of which luminosity $10^{27}$erg s$^{-1}$$<L_{X}<2\times 10^{31}$erg s$^{-1}$
with only upper limits for the other 73 systems.
Although the expected median X-ray luminosity is not so high, $L_{X}\le 10^{28.1}$erg s$^{-1}$, if both
detections and non-detections are all considered, a very
metal-poor binary, HD89499, with [Fe/H]$=-2.1$ emits
$L_X=1.3\times 10^{31}$erg s$^{-1}$, which is much larger than
the X-ray luminosity of the Sun, $L_{X,\odot}\sim 10^{27}$erg s$^{-1}$
\citep[e.g.,][]{gud04}.
Moreover, the temperature of the coronal plasma of HD89499 was found to be
very high, $T=2.6\times 10^7$K, from ASKA and ROSAT observations \citep{ft96}. 

Recently, X-ray was detected in the nearest Pop.II star with [Fe/H]$=-0.86$,
the Kapteyn's star, which is a single star having planets \citep{gui16}.
The detected X-ray luminosity is $L_X=(2.4-6.0)\times 10^{26}$erg s$^{-1}$.
Since the stellar mass is small, $M_{\star}=0.281M_{\odot}$, the bolometric
luminosity is only $L_{\rm bol}=0.012 L_{\odot}$. Therefore, the normalized X-ray
luminosity is $L_X/L_{\rm bol}\sim 10^{-5}$, which is quite large compared to
the solar value, $L_{X}/L_{\rm bol}\sim 10^{-7}-10^{-6}$, even though this star is
quite old with the age $\approx 11.5$ Gyr.

These observations indicate that at least some portions of metal-poor stars
exhibit high magnetic activity. Although the detailed properties of the
magnetic field are unknown and it is not well understood how the dynamo
process depends on metallicity, we expect that a sizable fraction of
metal-poor stars possesses magnetic field of which strength is comparable to
or larger than the solar value. Therefore, we think it is reasonable to
assume the magnetic flux tubes for metal-deficient stars introduced
in \S \ref{sec:LZMS}, whereas detailed wind properties also depend on the
filling factor, $f_0$, of open magnetic field \citep{suz06}.

We determined the velocity amplitude, $\delta v_0$, at the photosphere
from the convective flux by equation (\ref{eq:dv0}).
Although this estimate is expected to be
independent from metallicity, we should note that the transverse fluctuation
of magnetic flux tubes may depend on metallicity. \citet{mu02} analytically
evaluated the energy flux of vertically oriented magnetic flux tubes.
Their result shows that the energy flux of transverse fluctuations is
independent from metallicity for hotter stars with $T_{\rm eff} \gtrsim 6000$ K,
while it depends almost linearly on $Z$ for cooler stars with
$T_{\rm eff}\lesssim 4000$ K. Our adopted $\delta v_{0}$ is reasonable for
the cases with $M_{\star}\ge 0.7M_{\odot}$, while it may be overestimated
for the cases with $M_{\star}=0.6M_{\odot}$.

We mention the effect of stellar rotation, which have not been taken into
account in this paper. Stellar rotation affects the stellar atmosphere
and wind in two ways. First, if the rotation is fast, it directly affects the
velocity profile of the wind by the direct centrifugal acceleration \citep{wd67}.
Second, stellar rotation influences
the differential rotation in the surface convection zone \citep{bt02,hy11}, and
therefore, it probably affects the amplification of magnetic field there.
Low-mass stars lose their angular momentum throughout the pre-main
sequence and main sequence phases by
magnetized stellar winds \citep{wd67,hir97,vid09,pin11,mat12,jar13,rev15,joh15}.
In the present universe, long-lived low-mass Pop.III/II stars are probably
slow rotators as a result of the magnetic braking. Therefore, the
first effect of the centrifugal acceleration of the wind is unimportant for these stars.
%, and does not affect the results presented in this paper.
On the other hand, the second effect of the dynano action is probably
subect to the stellar rotation because the surface magnetic flux depends
on the rotation rate \citep{see18}. 

\subsection{Wave Dissipation}
\label{sec:wavedis}
Both the magnetic pressure associated with propagating \Alfvenic waves and
the gas pressure of the stellar coronae contribute to driving the stellar
winds from our low-mass Pop.II/III stars.  
The dissipation of the \Alfvenic waves is a key in driving the stellar winds
%from our low-mass Pop.II/III stars
because it controls the gas pressure
through the heating and the gradient of the \Alfvenic magnetic pressure
\citep{suz04,si06}.
In our simulations the injected \Alfvenic Poynting flux from the surface
mainly dissipates via nonlinear excitation of longitudinal waves
\citep{ks99,si05,nh06,nh07,mz10,vas11}.
Density fluctuations, which are regarded as longitudinal waves, are actually
detected in the solar wind by radio scintillation measurements using
AKATSUKI \citep{miy14}.
However, since our simulations are restricted to the simple 1D geometry, other
dissipation channels may be also important in more realistic situations.
We here briefly discuss wave dissipation mechanisms that we do not take into
account, referring to works for the solar corona and wind.

The injected \Alfvenic perturbation excites shear \Alfven waves with random
polarization in our simulations. However, if the effect of a magnetic flux tube
is directly considered, torsional \Alfven waves also have to be treated,
because the dissipation characters of shear and torsional modes are slightly
different \citep{nak00,vas12}.
In realistic 3D circumstances, \Alfven waves also dissipate via turbulent
cascade \citep{mat99,vv07,cra07,lio14,adh15,yan16,va16,tv17},
while the mode conversion to compressive waves will be suppressed because
propagating waves are not confined in a single flux tube.

It is still unclear how these processes modify
the dissipation rate in a quantitative sense. 
However, we can infer how the assumption of the 1D flux tubes affects the wave
dissipation from the comparison between simulations with
different dimensions. 2.5D MHD simulations by \citet{ms12,ms14} show that
the dissipation through the generation of compressive waves is suppressed,
compared to 1.5D simulations by \citet{si05,si06}. However, this suppression
is almost exactly compensated by the resistive dissipation by shearing motion
between neighboring magnetic field lines \citep{hp83}. As a result,
the total heating rates by the dissipation of \Alfven waves are not so
different at least between the 1.5D and 2.5D simulations. 
If this tendency can be extended to 3D simulations, our results of
the overall wave heating and the basic properties of the atmospheric structures
are expected to be reasonable. 

%対流層の深さは考慮していない

\section{Summary}
\label{sec:sum}
We investigated the structure of atmospheres and winds in
open magnetic field regions on low-mass stars with various metallicities.
We injected velocity fluctuations, of which the amplitude is evaluated
from the convective flux, from the location
at $T=T_{\rm eff}$ and solved MHD equations with radiative cooling and thermal
conduction in super-radially open one-dimensional magnetic flux tubes. 
By the dissipation of the \Alfvenic waves traveling from the photosphere 
hot coronae with $> 0.5\times 10^6$K are formed and coronal winds
stream out in all the simulated stars with $M_{\star}=(0.6-0.8)M_{\odot}$
and $Z=(0-1)Z_{\odot}$. 
However, the properties of the coronae and winds depend on metallicity.

Denser gas can be heated up to the coronal temperature for lower metallicity,
because the radiation cooling
%, which is proportional to $\rho^2$ in the optically thin gas,
is suppressed. As a result, the transition region that
separates the cool chromosphere and the hot corona is located at a lower
height with higher density. The coronal density of the stars with $Z\le 0.01
Z_{\odot}$ is 1-2 orders of magnitude higher than the coronal density of
the solar-metallicity star with the same stellar mass.

The difference between the density at the photosphere and the density in the
corona is smaller for lower-metallicity stars. Because density difference
determines the reflection of \Alfvenic waves, the smaller density contrast
leads to
a larger transmissivity of the \Alfvenic waves to the corona, which enhances
the heating in the corona. This enhanced heating, combined with the
suppressed radiation cooling, can explain the larger coronal density
in lower-metallicity stars. The coronal X-ray flux, which is proportional to
$\rho^2$, is also larger for lower metallicity, even though the cooling
efficiency, $\Lambda$ (erg cm$^{3}$s$^{-1}$), is smaller. 

We should note that this discussion is based on our simulations in open
magnetic field regions. In reality, X-rays are considered to come
dominantly from closed field regions on a stellar surface,
because denser plasma can be confined in closed loops.
%, referring to observations of the Sun (REFs).
Therefore, for quantitative estimates of X-rays, closed
magnetic loops need to be taken into account in our future studies.

The mass loss rate of the low-metallicity stars with $Z\le 0.01 Z_{\odot}$
is $(4.5-20)$ times larger than that of the solar-metallicity star with
the same mass, because of the larger coronal density. In terms of the
energetics, a larger fraction of the input \Alfvenic wave energy is transferred
to the kinetic energy of the stellar wind because of the suppression of the
wave reflection and the radiation loss.
%and the fraction that is reflected back to the photosphere.

It is interesting to note that the dependence of the mass loss rate on
metallicity is opposite to the trend for luminous stars. Stellar winds from
massive main sequence stars are driven by the radiation pressure acting
on metallic lines in the UV range \citep{ls70,cak75}.
The radiation pressure on dust grains plays
an important role in stellar winds from AGB stars \citep{bow88,ohn16}. 
These contributions are reduced for decreasing metallicity, and therefore,
the mass loss rate of these stars is smaller for lower metallicity
\citep{kud02,tas17}
%lower-metallicity massive stars \citep{kud02,mui12} and AGB stars
%\citep{tas17} is smaller than the higher-metallicity counterparts.
The main difference of less luminous stars we have studied in this paper
from these luminous stars is that the radiation acts as energy loss via
cooling in the coronal winds from low-luminosity stars, instead of the direct
momentum transfer by the radiation pressure. % in high-luminosity stars.

Our results also give an impact on the surface pollution of heavy elements
on low-mass Pop.III stars.  If the spherical Bondi-type accretion was assumed, 
low-mass Pop.III stars would not be observed as metal-free stars because of
the nonnegligible contribution of the metal pollution
\citep{yos81,kom15,she17}. However, if these stars had been driving stellar
winds of which the strength is comparable to that of the solar wind, the metal
pollution is negligible \citep{tan17}. Our results strongly support the latter
perspective, namely if low-mass Pop.III stars were formed at early epochs,
they could be detected as metal-free stars in the present-day universe.

%abstractの議論に加えて、$Z$小で$T_{eff}$大にも言及必要

We thank Hajime Susa and Shuta Tanaka for fruitful discussion.e
We also thank the referee for constructive comments. 
This work was supported in part by Grants-in-Aid for 
Scientific Research from the MEXT of Japan, 17H01105.

%\appendix
%\section{$Z$ dependence of $\rho_{\rm eff}$}
%\begin{figure}%[h]
%  \begin{center}
%    \includegraphics[width=0.46\textwidth]{../work/PhtsphonZ01.eps}
%  \end{center}
%  \caption{The data points indicate the density at $\tau_{\rm Ross}=0.3$ for
%    stars with $T_{\rm eff}=5750$ K and $\log g=4.5$ taken from ATLAS
%    \citep{kur79,ck03}, and the solid line denotes $D(Z)$.
%  \label{fig:Z-rhoph}}
%\end{figure}

%\bibliography{paper_lowmetalsw2017}

\end{document}